# AI Security Priorities

A Field-Wide Agenda

Gil Gekker[1], Rachel Steratore[2], Everett Smith[3,*], Asher Brass-Gershovich[2], Varun Gandhi[2], Nicole Nichols[4], Vijay Bolina[5], Buck Shlegeris[6], Lisa Einstein[7], Dan Lahav[1], Omer Nevo[1,†], Sella Nevo[2,†]

*[1] Irregular, [2] RAND, [3] Watertight AI, [4] Palo Alto Networks, [5] Prometheus, [6] Redwood Research, [7] The Hoover Institution*

## Abstract

As AI systems are rapidly integrated into critical economic, governmental, and national security functions, the gap between AI adoption and AI security readiness continues to widen. This paper presents a prioritized agenda for advancing AI security, informed by structured interviews with leaders across industry, government, and civil society, and refined through a multi-sector expert workshop. Participants identified and ranked the highest-importance and most cost-effective areas where progress could strengthen AI security – from protecting frontier AI systems and their underlying infrastructure to improving cybersecurity practices as AI reshapes the threat landscape. The resulting priorities are organized across four themes: establishing strategic foundations and policy frameworks; advancing public-private coordination and institutional infrastructure; advancing technical security engineering and assurance; and governing agentic AI under adversarial pressure. For each priority area, expert authors provide detailed analyses that define the problem, assess the current landscape, and identify actionable projects that stakeholders across sectors can pursue. The paper aims to serve as an initial practical foundation for coordinated investment and action across the AI security field. It is designed to serve both current practitioners and individuals and organizations looking to enter the field by identifying concrete, high-impact contributions suited to a range of strengths and capacities.

---

Corresponding author: gil@irregular.com

* Work conducted while at RAND.

† These authors contributed equally as senior authors.

## Acknowledgements

The authors are grateful to the many individuals who contributed to this project. We are especially thankful to Andrew Fasano, Jason Matheny, Joshua Saxe, Matt Malone, Miles Brundage, Phil Venables, and Tahira Mammen, as well as the various individuals from government, academia, and the frontier labs who shared valuable insights and helped shape the priorities reported here. Finally, we'd like to thank David Glikstein and Rosa Maria Torres for helping to shape the final version of this paper.

Acknowledged contributors do not necessarily endorse all statements in this paper.

# Disclosure Statement

## Conflicts of Interest

In accordance with RAND’s policies on conflicts of interest, the authors disclose the following financial and institutional relationships that are directly or topically related to the work described in this paper. The purpose of this disclosure is to provide readers with information about relationships and activities to fully understand this work.

**Gil Gekker, Dan Lahav, and Omer Nevo** are employed by Irregular (formerly Pattern Labs), an AI security lab that provides red-teaming, model evaluation, security assessment, and auditing services to frontier AI laboratories and governments. Several of this paper's recommendations — including the development of specifications for top-tier AI security red-teaming, the establishment of assessment and auditing protocols for AI companies and AI systems, the formation of public-private partnerships for AI security investment, and the expansion of AI security standards — are directly aligned with Irregular's commercial service offerings and could, if adopted, expand the market for services that Irregular provides. In addition, Irregular researchers co-designed the research methodology, including the interview protocol and workshop structure, and co-authored and reviewed priority area submissions for accuracy, consistency, and clarity. Dan Lahav and Omer Nevo are co-founders of Irregular and maintain equity in the company. Irregular may benefit financially from the adoption of this paper's recommendations.

**Everett Smith** is co-founder and Chief Operating Officer of Watertight AI, an early-stage startup that develops monitoring and oversight tools for autonomous AI agents, including systems designed to detect and correct reward hacking and misalignment in enterprise deployments. This paper's Chapter 4 recommendations — including the development of risk management frameworks for agentic AI, permission frameworks for AI agent interactions, and formal verification of security properties — are closely aligned with Watertight AI's product focus and could, if adopted, validate and expand the market for the types of agent oversight solutions Watertight AI provides.

## Independent Peer Review Process

To manage the conflicts of interest disclosed above, this paper was subjected to RAND’s rigorous quality assurance process, which requires that independent peer reviewers assess whether the paper's recommendations are well-supported by the evidence presented, and to evaluate whether any recommendations appear to have been influenced by the disclosed financial or institutional interests of the authors. Reviewers were required to disclose any relationships or activities that could bias their opinions of the manuscript. The reviewers have no financial interest in organizations or entities that could be impacted by the recommendations outlined in the paper.

Consistent with research demonstrating that informing peer reviewers of authors' conflicts of interest sharpens their scrutiny of methodology and conclusions, all reviewers were provided with the full text of this disclosure statement prior to conducting their review. Reviewers were specifically asked to evaluate: (1) whether the methodology — including the design of interviews, workshop structure, and scoring — contained features that could systematically favor outcomes beneficial to conflicted parties; and (2) whether the recommendations are proportionate to and supported by the evidence presented.

# Executive Summary

This paper synthesizes insights from more than 20 global artificial intelligence (AI) industry leaders, government officials, and experts spanning AI, cybersecurity, national security, technical research, policy, and governance who identified the highest-priority areas where progress could safeguard AI systems and advance frontier AI security. Expert input determined which priorities the field should pursue and how they rank. Individual sections were then developed by subject-matter experts drawing on their own knowledge and research, with lead authors providing editorial integration across the full report. If implemented, these actions could help tilt the advantage toward defenders over attackers, improve the security of AI systems, and build trust in AI.

AI's growing role in government operations, national security, and critical economic infrastructure underscores the urgent need for guidance on protecting these systems. Although AI offers many potential societal benefits, from improving decision-making to automating complex processes, its expanding capabilities introduce new risks, enabling bad actors to identify vulnerabilities and launch or scale cyberattacks. At the same time, advanced AI systems are becoming increasingly valuable targets, making them more likely to be subject to data poisoning, theft of industry secrets, or other forms of exploitation. Moreover, AI-generated code can accelerate the deployment of vulnerable software and harmful capabilities at scale. Despite these threats, the global AI ecosystem, including governments, private AI labs, major enterprises, and civil society, has yet to implement systematic, sustainable practices to advance AI security.

This paper addresses that gap by presenting the highest-priority areas across four themes that industry, government, and civil society leaders identified as having significant importance and high cost-effectiveness. These areas reflect the need for coordinated action across sectors and approaches, while also highlighting the many opportunities for individuals and organizations across disciplines to make meaningful contributions to advancing AI security.

For each area, we describe actionable projects that stakeholders across sectors and skillsets could pursue and identify concrete elements required for success. Collectively, these recommendations provide a practical foundation for durable, comprehensive AI security, helping stakeholders prioritize investments and coordinated actions.

## AI security priority area themes

These four themes organize the highest-priority areas for advancing frontier AI security. Together, they cover the strategic, institutional, technical, and operational dimensions needed both to protect AI systems as they become more capable, autonomous, and consequential, and to address the ways AI is reshaping the broader security landscape:

- **Establishing strategic foundations and policy frameworks:** As AI systems take on increasingly strategic roles and become targets for sophisticated adversaries, effective security requires more than the availability of technical solutions; even where fixes exist, barriers such as distribution, integration, organizational capacity, and cost can delay or prevent adoption. These priority areas focus on building coherent foundations for collective action: a national deterrence strategy to signal credible commitment to defending AI assets, technical standards that provide measurable security baselines, analytical frameworks connecting AI capabilities to downstream cybersecurity and societal outcomes (including how AI itself alters the threat landscape), a consolidated resource hub, and harmonized terminology to enable effective communication across the field.
- **Advancing public-private coordination and institutional infrastructure:** AI development is concentrated in a small number of private labs, yet these organizations face nation-state threats that exceed what any single company can address alone. At the same time, AI-enabled threats targeting critical infrastructure more broadly demand coordination that extends well beyond the labs themselves. These priority areas focus on building specific coordination mechanisms: government sharing of classified threat intelligence with labs, public investment in security measures that individual companies cannot justify commercially, personnel vetting adapted for access to sensitive AI capabilities, and dedicated channels for incident reporting and cross-organizational information sharing on both AI-specific and AI-enabled threats, and coordinated incident response and playbooks.
- **Advancing technical security engineering and assurance:** Frontier AI systems are both high-value targets for adversaries and sources of new security threats that legacy cybersecurity practices alone cannot fully address. As AI models become strategic assets and increasingly integrated into essential infrastructure, robust engineering and rigorous security assurance become increasingly important. These priority areas focus on developing, testing, and verifying technical protections for AI systems, as well as on harnessing AI to strengthen defensive capabilities, such as generating secure code and auditing systems at scale, while mitigating the risk that AI-generated software introduces.
- **Governing agentic AI under adversarial pressure:** As AI systems take on increasingly autonomous and agentic roles, they introduce risks that extend beyond traditional software compromise to include AI agents themselves acting against the systems and infrastructure they operate within. Traditional security measures and governance frameworks may no longer suffice to prevent novel forms of failure and misuse. These priority areas focus on securing AI in adversarial, high-stakes contexts through structured risk management, granular permissioning of agent actions, formal control mechanisms, and adversarial threat modeling and testing tailored for agentic systems, including scenarios in which AI agents themselves become the threat vector.

## 10 highest importance AI security priority areas

**Table S.1** presents the ten AI security priority areas that leaders in industry, government, and civil society identified as the most important. Each area represents a targeted focus where investment and coordinated action could yield the most significant improvements in AI security and resilience.

**Table S.1. 10 AI Security Priority Areas with the Highest Importance Scores**

| Ranking | Priority Area | Description | Theme |
|---|---|---|---|
| 1 | Develop a **national AI deterrence strategy** | Develop clear government policies to deter nation-state theft or sabotage of critical AI assets. | |
| 2 | Advance **information and incident sharing**** | Build dedicated, trusted systems for sharing threat intelligence, incident reports, and security lessons across AI labs, industry, and government. | |
| 3 | Develop **permission frameworks** for AI agent interactions | Develop structured frameworks for managing and auditing AI agent interactions with external environments, ensuring permissions are granular and governable by policy. | |
| 4 | Implement **confidential computing** to protect key AI system components | Secure sensitive AI components (e.g., model weights, code, data) during runtime using hardware-based enclaves and trusted execution environments. | |
| 5 | Research **secure code generation** to write more secure code | Developing AI techniques to produce more secure code and limit societal risks from potentially unsafe AI-generated software. | |
| 6 | Form **public-private partnerships** focused on AI security | Create frameworks for joint investment and collaboration between government and private AI labs, enabling security measures that exceed what any single organization could achieve alone. | |
| 7 | Develop **assessment and auditing protocols** for AI companies and systems | Develop protocols and tools to systematically audit and verify the security of AI companies and systems. | |
| 8 | Develop a **tamper-proof secure weight module** | Design and build specialized, tamper-resistant hardware modules to safeguard AI model weights against theft or misuse. | |
| 9 | Establish **formal verification of security properties** | Design and implement provable security mechanisms to prevent specific dangerous or unaligned AI actions, such as unauthorized exfiltration. | |
| 10 | Develop a **risk management framework for agentic AI** | Create lifecycle governance frameworks and continuous monitoring systems to manage, assess, and contain risks from increasingly autonomous AI agents. | |

**Theme 1:** Establishing strategic foundations & policy frameworks

**Theme 2:** Advancing public-private coordination & institutional infrastructure

**Theme 3:** Advancing technical security engineering & assurance

**Theme 4:** Mitigating risks of agentic AI under adversarial pressure

## 10 most cost-effective AI security priority areas

**Table S.2** presents the ten AI security priority areas that industry, government, and civil society leaders identified as most cost-effective. Each area represents a high-leverage opportunity where investment and coordinated action can deliver the greatest security benefit relative to effort and resources.

**Table S.2. 10 AI Security Priority Areas with the Highest Cost-Effectiveness Scores**

| Ranking | Priority Area | Description | Theme |
|---|---|---|---|
| 1 | Establish an **AI security resource hub** | Build a curated, open repository of trusted AI security tools, guidance, and educational materials. | |
| 2 | Create **specifications for top-tier AI security red-teaming** | Create standards and methodologies for rigorous, adversary-calibrated red teaming of AI systems and their supporting infrastructure. | |
| 3 | Develop **incident response frameworks and playbooks** | Develop coordinated frameworks and playbooks for responding to and reporting AI security incidents, tailored to the complexities of AI supply chains and deployment models. | |
| 4 | Develop **a risk management framework for agentic AI** | Create lifecycle governance frameworks and continuous monitoring systems to manage, assess, and contain risks from increasingly autonomous AI agents. | |
| 5 | Use **threat modeling to understand risks of AI self-attacks** | Develop comprehensive threat models and adversarial evaluation tools for scenarios where advanced AI systems could act as attackers against their own hosting infrastructure. | |
| 6 | Advance **information and incident sharing**** | Build dedicated, trusted systems for sharing threat intelligence, incident reports, and security lessons across AI labs, industry, and government. | |
| 7 | Establish **AI security standards** | Set concrete technical standards for securing AI systems and components. | |
| 8 | Understand how AI systems could impact society with **implications analysis** | Create frameworks linking specific AI capabilities to potential societal and cybersecurity outcomes. | |
| 9 | Clarify **language through a standard taxonomy** | Define and standardize key terms and concepts in AI security to improve communication across the field. | |
| 10 | Adapt **enhanced government background checks** for sensitive private personnel | Establish government-supported vetting and monitoring mechanisms for personnel with access to sensitive AI systems, addressing insider risks beyond standard commercial checks. | |

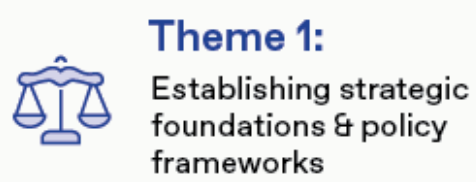
**Theme 1:** Establishing strategic foundations & policy frameworks

**Theme 2:** Advancing public-private coordination & institutional infrastructure

**Theme 3:** Advancing technical security engineering & assurance

**Theme 4:** Mitigating risks of agentic AI under adversarial pressure

**These priority areas are covered as a single section in Chapter 3: Information and incident sharing.

## First steps toward collaborative AI security

The priority areas and projects outlined in this paper underscore principles for industry, government, and civil society as they work toward universal AI security. **Meaningful progress will demand a variety of approaches and active participation from all sectors;** while no single actor can implement every initiative, engagement in any of the projects can make a significant contribution to global AI security. Each effort includes multiple project options of varying scopes, allowing participants to contribute according to their capacity and expertise. Moreover, **the projects presented here represent only a first draft of the AI security agenda**; as AI systems and the surrounding threat landscape continue to evolve, ongoing dialogue and flexible adaptation will be essential to meet emerging risks. With sustained, collaborative engagement, the AI ecosystem can move steadily toward a more secure and resilient future.

# Contents

# Chapter 1: AI Security at a Turning Point

As of early 2026, AI systems are being integrated into critical economic, governmental, and national security functions faster than the institutions responsible for securing them are adapting.[1,2] Organizations are deploying increasingly powerful AI models into software development, infrastructure management, decision support, and autonomous operations. This often occurs without established methods to assess, mitigate, or coordinate responses to the resulting security risks. The gap between AI adoption and AI security readiness is widening at a moment when the consequences of failure are growing more severe.

Evidence of this widening gap is clear. According to the World Economic Forum's *2025 Global Cybersecurity Outlook*,[3] 66% of organizations expect AI to significantly affect cybersecurity in the coming year, yet only 37% report having processes to evaluate the security of AI tools before deployment. Meanwhile, adversaries are not waiting. The *2024 Annual Threat Assessment*[4] identifies AI as a priority target for foreign intelligence services, and AI experts in this study (see Appendix A) consistently reported active targeting of U.S. AI capabilities by nation-state actors. These adversaries coordinate across campaigns, share tools and tradecraft, and pursue long-term strategic objectives;[5] in contrast, defenders often operate in isolation.

At the same time, AI is reshaping the offense-defense balance. AI-enabled tools are accelerating vulnerability discovery, scaling social engineering, and compressing attack timelines in ways that strain human-centric security processes.[6] The rapid adoption of AI-generated code compounds the problem: as AI systems increasingly write software at scale, they may also propagate insecure patterns and latent vulnerabilities faster than traditional review and remediation processes can keep pace. These dynamics create systemic risk rather than isolated failure modes, raising the stakes for both coordinated, field-level responses and contributions by individuals and organizations across industry, government, and civil society.

## A field transformed

The meaning of "AI security" has shifted dramatically over the past decade. Early work in the field focused primarily on adversarial machine learning: crafting inputs that fool classifiers, studying robustness to perturbations, and defending against data poisoning. These efforts were foundational, but they assumed a world in which machine learning models were narrow components embedded within larger systems, performing bounded tasks such as image recognition or fraud detection. As AI systems grew more capable and more deeply integrated into critical functions, the field's scope expanded well beyond adversarial robustness.

The emergence of large language models (LLMs) and increasingly general-purpose AI systems, beginning with GPT-3 in 2020 and accelerating through 2023-2026, has been a primary driver of this shift. AI systems have evolved from specialized components into platforms that write code, reason over complex domains, operate with growing autonomy, and underpin core organizational functions. In response, "AI security" has come to encompass the security challenges, capabilities, and practices that arise at the intersection of cybersecurity and AI. This includes safeguarding the critical assets of AI systems such as model weights, evaluating whether models increase the marginal risk of cyberattacks, developing security and compliance standards for AI, using AI to strengthen defensive cyber tools, and applying information security techniques to contain and control potentially malicious AI agents. This is the definition of AI security we adopt throughout this paper. Some recent usage extends the term further to encompass AI's broader role in national security strategy; while this paper touches on strategic dimensions where they intersect with cybersecurity, that broader framing is not our primary scope. Throughout this paper, we use the term frontier AI to refer to the most capable general-purpose AI systems at or near the leading edge of current development, which pose the greatest security stakes and are the primary focus of this agenda.

## Emerging security challenges

As AI has become a tool for both attackers and defenders, its evolution has introduced several new AI security challenges:

**First, AI systems have become high-value targets.** When a model supported a single application, compromise affected a limited scope. When a foundation model underlies customer service, software development, data analysis, and decision support, failures can cascade across an entire organization or ecosystem. Model weights now represent billions of dollars in R&D investment and confer significant economic advantages; many expect them to become sources of meaningful geopolitical advantage, making them priority targets for nation-state intelligence services. AI security is no longer only a technical concern; it is a matter of strategic competition and national interest.

**Second, AI is transforming the cyber threat landscape.** AI is becoming a powerful defensive tool, automating code review, detecting anomalies, and enabling security operations at a pace and scale that human analysts alone cannot sustain. Understanding how AI shifts the offense-defense balance and tilting that balance toward defenders by ensuring that defensive applications outpace offensive ones, is a central challenge for the field.

**Third, AI systems are becoming autonomous actors.** As AI agents are deployed with real access to infrastructure, tools, and decision-making authority, they introduce risks that extend beyond traditional software vulnerabilities. An agentic system that can plan, execute multi-step tasks, and modify its environment must be governed and constrained through architectural and institutional controls, not just model-level safeguards. These risks extend cybersecurity beyond preventing software compromise to the more complex challenge of managing the behavior and authority of autonomous AI agents.

These shifts mean that AI security is not simply an extension of traditional cybersecurity or software assurance. It is a field that now spans strategy, policy, institutions, engineering, and governance, and that requires coordinated action across these layers. At the same time, it should build on the substantial body of existing information security infrastructure, frameworks, and hard-won operational insights rather than starting from scratch. The goal is to adapt and extend proven approaches where they apply, and to develop new capabilities only where AI-specific risks genuinely demand them.

## Approach and results: cost-effective and important priority areas for AI security

The scale and urgency of these challenges demand a clearly prioritized agenda. The work ahead is vast, resources are finite, and the window to address the most consequential risks is narrowing. This paper identifies where targeted investment and action could most meaningfully improve the state of AI security. To do so, we convened more than 20 global experts from across the AI security field, spanning AI, cybersecurity, national security, technical research, policy, and governance, and asked them to identify the ten most important and ten most cost-effective priorities given the current threat landscape.

To identify and prioritize these challenges, we engaged experts through a multi-stage process. We first conducted scoping interviews with senior, global experts in AI governance, cybersecurity, and AI development, then convened a structured workshop with 14 participants from industry, government, and civil society to refine and score the most promising priority areas. Seven additional experts assessed the practical difficulty of implementation for each area. We combined these inputs into a cost-effectiveness calculation and selected the top priorities by both importance and cost-effectiveness, merging closely related priority areas where appropriate.[1] The expert process determined which priority areas to pursue and how they ranked relative to one another. The detailed analysis within each priority area (defining the problem, explaining why it matters, proposing projects, and outlining elements of success) was then developed by individual expert authors matched to areas based on their subject-matter expertise. These authors drew on their own knowledge, research, and professional experience, not solely on what was discussed during the workshop. Lead authors provided editorial integration, framing, and coherence across sections. The result is a synthesis that combines structured collective prioritization with deep individual expertise. The prioritization process produced a set of clear, actionable leverage points where effort applied now could materially reduce risk. The two organizing metrics are:

- **Importance:** The average expert rating of how significant each priority area is for advancing AI security.
- **Cost-effectiveness:** A composite metric reflecting importance relative to the effort required and the degree to which the area is already being addressed. In this context, "cost" refers not to financial expenditure alone but to the broader effort involved, including organizational complexity, coordination demands, and implementation difficulty. Areas that score high on cost-effectiveness represent high-leverage opportunities where additional investment is likely to yield outsized returns relative to the effort required.

[1] Full methodological detail, including recruitment procedures, scoring methods, and limitations, is in Appendix A.

For each priority area, expert authors of this report provide detailed analyses that define the problem, explain why it matters, propose concrete projects, and outline what success looks like. Whether the reader is a researcher seeking open problems, a policymaker evaluating institutional gaps, a funder shaping a portfolio, or a practitioner looking to contribute to this emerging field, each section is designed to stand on its own as a grounded starting point for action.

Tables 1 and 2 present the priority areas identified as having the highest importance and highest cost-effectiveness, respectively.

**Table 1. 10 AI Security Priority Areas with the Highest Importance Scores**

| Ranking | Priority Area | Description | Theme |
|---|---|---|---|
| 1 | Develop a **national AI deterrence strategy** | Develop clear government policies to deter nation-state theft or sabotage of critical AI assets. | |
| 2 | Advance **information and incident sharing**** | Build dedicated, trusted systems for sharing threat intelligence, incident reports, and security lessons across AI labs, industry, and government. | |
| 3 | Develop **permission frameworks** for AI agent interactions | Develop structured frameworks for managing and auditing AI agent interactions with external environments, ensuring permissions are granular and governable by policy. | |
| 4 | Implement **confidential computing** to protect key AI system components | Secure sensitive AI components (e.g., model weights, code, data) during runtime using hardware-based enclaves and trusted execution environments. | |
| 5 | Research **secure code generation** to write more secure code | Developing AI techniques to produce more secure code and limit societal risks from potentially unsafe AI-generated software. | |
| 6 | Form **public-private partnerships** focused on AI security | Create frameworks for joint investment and collaboration between government and private AI labs, enabling security measures that exceed what any single organization could achieve alone. | |
| 7 | Develop **assessment and auditing protocols** for AI companies and systems | Develop protocols and tools to systematically audit and verify the security of AI companies and systems. | |
| 8 | Develop a **tamper-proof secure weight module** | Design and build specialized, tamper-resistant hardware modules to safeguard AI model weights against theft or misuse. | |
| 9 | Establish **formal verification of security properties** | Design and implement provable security mechanisms to prevent specific dangerous or unaligned AI actions, such as unauthorized exfiltration. | |
| 10 | Develop a **risk management framework for agentic AI** | Create lifecycle governance frameworks and continuous monitoring systems to manage, assess, and contain risks from increasingly autonomous AI agents. | |

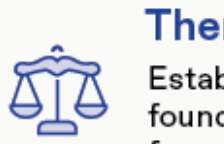

**Theme 1:** Establishing strategic foundations & policy frameworks

**Theme 2:** Advancing public-private coordination & institutional infrastructure

**Theme 3:** Advancing technical security engineering & assurance

**Theme 4:** Mitigating risks of agentic AI under adversarial pressure

**Table 2. 10 AI Security Priority Areas with the Highest Cost-Effectiveness Scores**

| Ranking | Priority Area | Description | Theme |
|---|---|---|---|
| 1 | Establish an **AI security resource hub** | Build a curated, open repository of trusted AI security tools, guidance, and educational materials. | |
| 2 | Create **specifications for top-tier AI security red-teaming** | Create standards and methodologies for rigorous, adversary-calibrated red teaming of AI systems and their supporting infrastructure. | |
| 3 | Develop **incident response frameworks and playbooks** | Develop coordinated frameworks and playbooks for responding to and reporting AI security incidents, tailored to the complexities of AI supply chains and deployment models. | |
| 4 | Develop **a risk management framework for agentic AI** | Create lifecycle governance frameworks and continuous monitoring systems to manage, assess, and contain risks from increasingly autonomous AI agents. | |
| 5 | Use **threat modeling to understand risks of AI self-attacks** | Develop comprehensive threat models and adversarial evaluation tools for scenarios where advanced AI systems could act as attackers against their own hosting infrastructure. | |
| 6 | Advance **information and incident sharing**** | Build dedicated, trusted systems for sharing threat intelligence, incident reports, and security lessons across AI labs, industry, and government. | |
| 7 | Establish **AI security standards** | Set concrete technical standards for securing AI systems and components. | |
| 8 | Understand how AI systems could impact society with **implications analysis** | Create frameworks linking specific AI capabilities to potential societal and cybersecurity outcomes. | |
| 9 | Clarify **language through a standard taxonomy** | Define and standardize key terms and concepts in AI security to improve communication across the field. | |
| 10 | Adapt **enhanced government background checks** for sensitive private personnel | Establish government-supported vetting and monitoring mechanisms for personnel with access to sensitive AI systems, addressing insider risks beyond standard commercial checks. | |

**Theme 1:** Establishing strategic foundations & policy frameworks

**Theme 2:** Advancing public-private coordination & institutional infrastructure

**Theme 3:** Advancing technical security engineering & assurance

**Theme 4:** Mitigating risks of agentic AI under adversarial pressure

**These priority areas are covered as a single section in Chapter 3: Information and incident sharing.

Across the full set of priorities, several patterns emerge. The most cost-effective opportunities tend to be foundational: building shared resources, developing assessment protocols, and establishing incident response practices. These are efforts where relatively modest investment could unlock outsized returns by giving the field common tools and frameworks it currently lacks. The most important priorities often involve larger-scale institutional or governmental action, such as national deterrence strategy, intelligence sharing with AI labs, and confidential computing research. Many of the most important priorities ranked among the most costly and complex to execute, underscoring the need for sustained, long-term commitment alongside near-term wins. The priorities identified here also represent a first draft of the AI security agenda rather than the final word; as AI systems and the surrounding threat landscape evolve, the field will need ongoing dialogue and flexible adaptation to keep pace.

## Report roadmap

The report is organized into four chapters. Each chapter addresses a distinct dimension of AI security, and together they aim to close the gap between the pace of AI adoption and the maturity of the security practices, institutions, and governance frameworks the field requires.

**Chapter 2: Strategic foundations and policy frameworks** examines the strategic, conceptual, and analytical infrastructure required for AI security to function as a coherent field. Topics include United States Government (USG) AI national deterrence strategy, technical standards, implications analysis, and shared language, with a focus on frameworks that enable coordinated action across technical and policy domains. Priority areas covered:

- Develop a national AI deterrence strategy
- Establish AI security standards
- Understand how AI systems could impact society with implications analysis
- Establish an AI security resource hub
- Clarify language through a standard taxonomy

**Chapter 3: Public-private coordination and institutional infrastructure** focuses on the mechanisms needed for effective collaboration between government and industry. Topics include public-private security investments, personnel risk management, intelligence sharing, information-sharing organizations, and incident response coordination, with attention to the pressures created by global competition and the concentration of critical AI capabilities in a small number of organizations. Priority areas covered:

- Form public-private partnerships focused on AI security (including government sharing of classified threat intelligence with AI labs)
- Adapt enhanced government background checks for sensitive AI lab personnel
- Advance information and incident sharing
- Develop incident response frameworks and playbooks

**Chapter 4: Technical security engineering and assurance** turns to the engineering layer: how AI systems can be secured in practice and how security claims can be meaningfully verified. Topics span confidential computing, secure code generation, security posture assessment, protection of model weights, and rigorous red-teaming, with an emphasis on what can be built and verified with current technology and where further research is needed to close critical gaps. Priority areas covered:

- Implement confidential computing to protect key AI system components
- Research secure code generation to write more secure code
- Develop assessment and auditing protocols for AI companies and AI systems
- Develop a tamper-proof secure weight module for AI model weights
- Create specifications for top-tier AI security red-teaming

**Chapter 5: Governing agentic AI under adversarial pressure** explores the emerging challenge of securing increasingly autonomous AI systems. Topics include system-level governance, control mechanisms, and adversarial threat models for agentic AI deployed with real authority and access to critical infrastructure, examining how architectural and governance approaches outside the model itself can constrain autonomous systems in practice. Priority areas covered:

- Develop permission frameworks for AI agent interactions
- Establish formal verification of security properties
- Develop a risk management framework for agentic AI
- Use threat modeling to understand risks of AI attacking its host infrastructure

# Chapter 2: Strategic foundations and policy frameworks

As AI emerges as a strategic asset with implications for national security and societal well-being, securing it against sophisticated adversaries will require more than technical solutions alone. **This chapter presents five high-level priority areas that each involve broad societal impact, including policy measures and cross-sector coordination efforts**. Together, these priority areas aim to strengthen AI security across government, industry, and civil society by clarifying strategic priorities, standardizing practices, and consolidating resources and knowledge. By presenting each theme's significance, suggested projects, and elements of success, this chapter lays the groundwork for subsequent technical work, policy development, and cross-sector coordination.

The following priority areas illustrate where targeted actions can strengthen AI security:

1. **National deterrence strategies** establish how states signal credible commitments to defend AI systems from adversarial nations.
2. **Technical standards** provide concrete, measurable security requirements that organizations can implement and auditors can verify.
3. **Implications analysis,** as used in this report, refers to systematic assessment of how AI capabilities could affect society across domains such as critical infrastructure, offensive cyber operations, and national security. It creates methods for connecting capabilities to their downstream effects, enabling prioritization of security investments.
4. **AI security resource hub** lowers the bar for organizations to strengthen their defensive posture by consolidating fragmented knowledge into a single, accessible repository.
5. **Taxonomy and language clarification** ensure that different communities share sufficient conceptual common ground when discussing "AI security."

## Develop a national AI deterrence strategy

Importance #10

### What is this?

A clear strategic posture could help deter nation-state theft, sabotage, or coercion targeting U.S. AI systems, infrastructure, and intellectual property, filling a significant gap in U.S. strategy. This would include establishing credible commitments to specific responses (such as economic sanctions and diplomatic consequences) in the event of model weight theft, attacks on AI infrastructure, or other hostile actions against the U.S. AI ecosystem. Such a strategy would extend existing U.S. cyber and technology deterrence frameworks, building on the National Cyber Strategy[7] and Department of War (DoW) cyber deterrence posture.[8] Such a strategy would include two distinct components: an internal policy framework (detailed, nuanced, and largely classified) that guides interagency decision-making, and a public declaratory posture (concise and deliberately less specific) designed to shape adversary calculations.

### Why this matters

Several factors make developing such a framework urgent for protecting U.S. AI systems:

- **Multiple governments have identified AI leadership as a national priority with strategic implications.** The National Security Memorandum on AI (October 2024) declared that protecting U.S. AI intellectual property and infrastructure from foreign intelligence threats is official U.S. policy, designating it a "top-tier intelligence priority."[9] The White House AI Action Plan (July 2025) similarly identifies maintaining U.S. AI leadership as a core strategic goal and warns against adversaries free-riding on American innovation.[10] As AI capabilities advance, several analysts have argued that frontier AI systems could become strategically significant assets, with implications for military capability, economic competitiveness, and intelligence operations.[11] The dual-use nature of advanced AI means that capabilities developed for civilian purposes could be repurposed for military or intelligence applications,[12] raising the national security stakes of theft or sabotage.
- **U.S. adversaries might not perceive consequences from attacking U.S. AI systems.** Without a clear deterrence framework, the United States risks a strategic environment where either adversaries face minimal consequences for aggressive actions against U.S. AI assets, or the United States responds to incidents reactively, paying the costs of escalation without having gained the preventive benefits of a declared deterrence posture.
- **Frontier AI model weights represent a growing nation-state interest:** They represent billions of dollars in R&D investment and potential strategic advantages. Intelligence assessments consistently highlight that major adversaries are actively targeting U.S. AI capabilities through cyber operations, human intelligence, and supply chain compromise. The 2024 Annual Threat Assessment identifies AI as a priority target for foreign intelligence services,[4] a trend that only intensified through 2025 as frontier models grew more capable and commercially valuable - and one likely to deepen further as AI capabilities become more consequential.

- **Private ownership of frontier AI systems presents an ambiguous U.S. posture:** Unlike traditional defense assets, frontier AI systems sit primarily within private companies. There are no clear government commitments regarding the protection of these assets or consequences for attacks against them. Adversaries may reasonably calculate that stealing AI assets or sabotaging AI infrastructure carries acceptable risk, since the U.S. has not articulated what responses such actions would trigger. This gap between the strategic importance of AI assets and the absence of a declared posture creates an environment that invites probing and escalation.
- **The United States faces attribution complexities:** AI-related cyber operations and IP theft present attribution challenges that any deterrence framework must address. The U.S. government generally succeeds in attributing major cyber operations to their source, often identifying the specific threat actor. However, establishing publicly credible proof is harder, which gives adversaries a degree of plausible deniability.[13] A deterrence strategy must establish how the U.S. will act on its attribution judgments, including in cases where adversaries dispute responsibility.
- **Establishing a deterrence framework is feasible and timely**: Enabling policy development, interagency coordination, and strategic communication can contribute to credible deterrence without requiring new technical capabilities. Acting proactively - before major incidents occur - can increase credibility and allow measured interventions rather than risky responses.

## Suggested projects

Establishing a national AI deterrence posture is fundamentally an interagency policy challenge. The U.S. government has extensive experience developing deterrence frameworks in other domains, but no office currently owns the process of extending these to AI-specific threats. The projects below move from foundational research that a broad range of policy researchers can undertake to the core interagency work that only senior government officials can drive. Projects include:

- **Graduated response principles.** Independent policy researchers could develop principles for how different categories of hostile action (model weight theft, infrastructure sabotage, supply chain attacks, coercion of AI personnel) relate to different levels of U.S. response. In practice, U.S. deterrence postures maintain some ambiguity about exact responses to preserve flexibility. The goal is not a rigid public mapping of actions to consequences, but rather a coherent framework that ensures proportionate decision-making that can inform the eventual interagency process and shape adversary calculations. Such research, conducted by think tanks or academic institutions, would lower the activation energy for the government to adopt a formal posture by providing an analytically grounded starting point.
- **Legal authority mapping to adversary dynamics.** Three research areas would significantly strengthen any future policy process by providing inputs that do not currently exist in consolidated form: an assessment of existing U.S. legal authorities for responding to AI-related hostile actions and gaps that may require new legislation or executive action; an analysis of how key adversaries are likely to perceive different U.S. postures, including escalation dynamics; and an examination of how AI deterrence interacts with broader strategic competition. Consolidating these inputs would reduce the analytical burden on the interagency process and make it more tractable when initiated. For those entering this area, RAND's recent work on AGI and international security provides a useful starting point, including analysis of denial, punishment, and governance approaches to protecting AI development from state interference.[14]

- **Driving the interagency policy process to produce an internal strategy document.** The central project is initiating and sustaining the interagency coordination needed to produce a formal U.S. deterrence posture for AI assets. The primary output is an internal strategy document that establishes U.S. policy on protecting AI assets and the consequences for hostile action against them; the public-facing declaratory elements would be derived from this internal framework. This means engaging the relevant offices across the NSC, State Department, Treasury, Commerce, Defense, and the Intelligence Community; building the case for why this posture is needed now; developing the specific policy language; and shepherding it through the approval process. This work is analogous to the process that produced the National Cybersecurity Strategy's declaratory elements. This effort could include establishing a public-private coordination mechanism, since the assets being protected reside primarily in private companies. This work can realistically only be led by officials in the relevant government positions, but those building expertise in AI security policy, today could be well-positioned for such roles in the future.

## Elements of success

A successful AI deterrence framework must navigate several tensions. The goal of this priority area is not to establish perfect deterrence, but to shift adversary cost-benefit calculations enough to reduce the frequency and severity of hostile actions against U.S. AI assets. Even partial success would meaningfully improve the security environment for American AI development. The following are not exhaustive requirements but represent important dimensions where careful design choices are needed. Successful projects will:

- **Establish credibility and remain flexible:** Deterrence requires that adversaries believe the United States will follow through on its commitments. However, overly rigid public commitments may constrain U.S. options or create escalation risks. Frameworks should balance clear, believable signals of resolve with sufficient flexibility to respond proportionately to scenarios that do not fit neatly into pre-defined categories, such as maintaining credibility when attribution is contested.
- **Engage the private sector:** Since key AI assets reside in private companies, deterrence frameworks should address how government and industry coordinate. This includes defining company obligations, government commitments to assist companies facing state-directed threats, and management of potential tensions between security requirements and commercial interests (including international operations).
- **Maintain proportionality and avoid escalation:** Responses should be severe enough to impose real costs on adversaries but proportionate enough to maintain international legitimacy and avoid spiraling escalation. Achieving this balance requires careful calibration, which can vary depending on the adversary and nature of the hostile action.

- **Balance transparent signaling and discretion:** As noted above, the internal strategy and public declaration serve different functions and audiences. Effective deterrence requires carefully balancing what is communicated publicly and what is retained privately: clearly signaling red lines can shape how adversaries behave, while keeping response options and intelligence sources undisclosed preserves flexibility. A successful strategy thoughtfully balances these considerations.
- **Foster interagency coordination:** AI deterrence involves multiple agencies with different equities and authorities. Projects should leverage clear coordination mechanisms and work to integrate existing cyber deterrence and broader national security frameworks.
- **Manage the security dilemma:** Projects should carefully calibrate public messaging to balance deterrence and perceived vulnerability. Explicitly designating AI assets as strategic targets could unintentionally signal both their value and their vulnerability, potentially incentivizing adversary action. Deterrence frameworks in other domains have faced this same dynamic, underscoring the need to signal resolve without exposing vulnerability.

# Establish AI security standards

Cost-effectiveness #7

## What is this?

AI security standards define how to protect AI systems from both unintentional failures and intentional adversarial attacks. Developing these standards requires addressing three fundamental questions: (1) what critical components must be secured, (2) what threats they face, and (3) what constitutes adequate protection for each component in context. The first two questions form the inseparable design criteria, and the third represents defensive research challenges that are addressed in the suggested projects and elements of success sections below.

AI systems present unique challenges that go beyond traditional software security. These include protecting model weights, ensuring the integrity of training data, defending against inference-time attacks, addressing novel AI-native vulnerabilities, such as jailbreaks and distillation attacks, and mitigating emerging risks like multi-agent collusion.

As AI systems become increasingly agentic, standards should additionally address other fundamentally new challenges such as task alignment, execution boundaries, and inter-agent communication.[15] Effective standards translate these complex requirements into adoptable, measurable, and technically grounded specifications that organizations can consistently implement across diverse deployment contexts. Because AI capabilities are rapidly evolving, standards should remain closely connected to the evolving threat landscape and security practices to stay adoptable and relevant.

## Why this matters

The absence of agreed-upon AI security standards creates systemic vulnerabilities at a moment of accelerated AI innovation and adoption. Key reasons that establishing standards matters include:

- **AI autonomy increases potential risks.** AI systems are increasingly deployed as autonomous agents that take actions in real-world environments: executing code, managing infrastructure, handling sensitive data, and interacting with external services. As these systems grow in autonomy and capability, the attack surface expands in ways that lack established security boundaries. Without standards that define secure execution constraints, permissible agent behaviors, and trust boundaries, organizations have no common basis for determining what "secure" means for an agentic system. This challenge could intensify as agents become more interconnected and multi-agent deployments grow.
- **Inconsistent protections leave gaps.** Without agreed-upon standards, each organization deploying AI in sensitive contexts - healthcare, finance, critical infrastructure, national security - must independently assess AI security. This leads to inconsistent protections, duplicated effort, and gaps that adversaries can exploit. Standardized security frameworks could help establish common baselines that enable responsible AI deployment at scale while reducing the overhead of bespoke security assessments.

- **High-value systems require proactive safeguards.** As AI systems grow in economic and strategic value, the stakes of security failures are substantial. Model weights representing billions of dollars in development investment are high-value targets for exfiltration by insiders or sophisticated external attackers, and AI-dependent infrastructure will increasingly factor into geopolitical competition. Establishing robust AI security standards now - before these systems become deeply embedded in critical functions - could help organizations protect their investments and contribute to long-term economic and political stability.

## Suggested projects

Several categories of work could advance AI security standards development, including:

- **Component-security matrix development:** A foundational project would systematically map security objectives against AI system components. A suggested starting point for security objectives includes redundancy, supervision, isolation, and identity/permissions/attestation, while the following is an example list of key AI system components: hardware, containers, networking components, models, and data. Taking these parameters and building upon them through this matrix approach can help standards developers identify which controls apply where, and help ensure comprehensive coverage. For instance, the matrix might clarify that hardware redundancy may be impractical except for critical infrastructure, while container isolation should apply universally across microservices. The project could produce not just the matrix itself but guidance on how organizations should interpret and apply it across different deployment contexts, risk levels, and operational requirements. Industry working groups could validate the matrix against real-world AI architectures to ensure practical applicability.
- **Working memory safety standards:** LLMs and AI agents depend on various forms of working memory (context windows, conversation history, retrieved documents) that create unique attack surfaces. New standards could help detect multi-turn prompt injections, jailbreaking attempts, and data exfiltration through context manipulation. This project could specify detection requirements, acceptable latency bounds for production systems, evaluation methodologies including variable sliding windows within prompt streams, and taxonomies of multi-turn attack strategies. This work could enable vendors to build interoperable monitoring solutions and organizations to establish consistent protection levels.
- **Training phase security specifications:** Model weights and training data require protection throughout the development lifecycle without impacting researcher productivity or training efficiency. Standards development could address secure attestation and permissioning systems that can operate with negligible throughput impact while providing end-to-end provable integrity. Specifications could include acceptable latency targets that may vary across throughput stages, with stricter requirements proportional to bottleneck severity and security risk at each point in the data flow. This work would require collaboration between security engineers and ML infrastructure teams to ensure standards remain practical for the unique demands of large-scale model training.

- **Security level standards for model and algorithmic insight protection:** As organizations adopt tiered security frameworks - such as those defining security levels (e.g., SL-3 – SL-5[16]) for protecting model weights and related assets - standards could help specify what each level concretely requires in terms of technical controls, access policies, and monitoring. Parallel standards could address protection of algorithmic insights[17] (training methodologies, architectural innovations, evaluation results) that might carry significant strategic value beyond model weights. This work could define clear criteria for when each security level applies, what controls satisfy it, and how organizations can verify compliance.

## Elements of success

Developing effective AI security standards will depend on practical implementability, comprehensive coverage, and alignment with real-world threats. Successful projects will:

- **Ensure adoptability across diverse organizations and AI system types:** Standards should include clear guidance on which requirements apply in which contexts and why. Even within a single enterprise, ambiguous standards lead to inconsistent interpretation across teams, creating confusion and security gaps. Successful standards include clear ownership for data, evaluations, and incident response, and applicability across system types and deployment contexts. They should further include practical implementation pathways that account for varying organizational capabilities. Standards bodies should engage implementers to identify and address adoptability barriers before publication, likely requiring such bodies include industry leaders and experts.
- **Ensure standards are rooted in technical grounding:** AI security standards should be rooted in the actual technical characteristics of the systems they protect. Wherever possible, requirements should derive directly from measurable operational conditions rather than abstracted or secondary measures that introduce ambiguity. For example, standards for model weight isolation should specify technical controls (encryption, access controls, network segmentation) with measurable properties rather than vague requirements to "adequately protect" weights. Similarly, monitoring standards should specify detection latencies, false positive rates, and coverage requirements that can be objectively verified. Technical grounding ensures standards remain meaningful as AI systems evolve and prevents compliance that does not deliver meaningful security.
- **Adopt a holistic scope:** Standards should address the full AI system - spanning hardware, software, supply chain, conventional cyberattack vectors, and AI-specific attack vectors - as strong defenses in one part of the system can be easily circumvented elsewhere. Standards development should explicitly address interfaces between components and phases of the AI lifecycle, ensuring that security measures compose effectively rather than creating false confidence through isolated controls. Standards should further account for both safety and security - addressing the potential for unintentional harms arising from system failures or errors, as well as intentional threats from adversaries seeking to exploit, manipulate, or degrade systems.
- **Adopt realistic threat modeling:** Standards should reflect the actual threat landscape, incorporating attacker profiles and risks specific to the system’s value and exposure.

What security researchers observe in the public domain and what is publicly disclosed represent only a fraction of actual attack activity, particularly for high-value targets like frontier AI systems. Standards development should incorporate threat intelligence from organizations with visibility into sophisticated attacks, including government agencies and large enterprises that see attack volumes and techniques unavailable to most researchers. This realistic threat modeling could help prioritize which security measures matter most and ensure standards address the attacks organizations actually face, not just those that appear in published research.

- **Refine iteratively for future uncertainty:** Security standards should evolve as AI capabilities and attack techniques evolve. There is a shrinking margin between the advancement of a capability and the time available to appropriately adapt. Standards development timelines should account for how quickly new AI capabilities are proliferating, so that security requirements are in place before vulnerable systems are widely deployed. Standards development processes should include mechanisms for regular review and ongoing engagement between standards bodies, AI developers, security researchers, and operational teams who encounter novel attacks. Standards should also specify how organizations should handle emerging threats that fall outside current requirements, providing frameworks for risk-based decision-making rather than rigid compliance checklists that become obsolete as the threat landscape shifts.

# Understand how AI systems could impact society with implications analysis

Cost-effectiveness #8

## What is this?

Analytical frameworks and methodologies could help trace how specific AI system-enabled offensive cyber capabilities - from vulnerability exploitation techniques to AI-augmented attack tools - could influence downstream societal effects. These analysis tools would articulate causal hypotheses and influence pathways that link capability development and operational deployment to economic, political, security, and social outcomes. Because cyber and societal systems are complex and interconnected, the relationships between them might often indicate trends or associations rather than direct cause-and-effect, requiring analysts to rely on counterfactual reasoning and structured comparisons rather than direct observation. Implications analysis frameworks could help policymakers, researchers, and security professionals anticipate plausible consequences before capabilities proliferate, rather than reacting only after harms materialize.

## Why this matters

Several factors underscore the importance of analyzing how AI-enabled cyber capabilities translate into real-world consequences:

- **Implications analysis can help identify non-obvious relationships between capabilities and outcomes:** Societal impact often depends on factors far removed from technical specifications. For example, a vulnerability exploitation technique might remain dormant for years, then enable catastrophic attacks when paired with changes in target infrastructure, attacker motivations, or geopolitical context. Past events, such as the 2017 WannaCry ransomware attack, demonstrate how non-obvious factors, such as North Korean strategic objectives, National Health Service IT architecture, and the maturation of Bitcoin, could enable a single leaked capability to cascade into global disruption. Implications analysis could help policymakers identify these influencing factors.
- **Implications analysis can help clarify how AI is transforming the capability landscape and might generate societal consequences:** AI-augmented offensive tools are emerging across the attack lifecycle - from automated vulnerability discovery to adaptive social engineering to autonomous lateral movement. These capabilities do more than amplify existing attack patterns; they can create new risks entirely. For example, recent analysis suggests AI systems are beginning to demonstrate emergent offensive cyber capabilities - discovering and exploiting novel vulnerabilities, adapting attack strategies in response to defenses, and exhibiting sophisticated behavior without explicit training for these tasks. Implication analysis can help map these AI capabilities to likely societal outcomes and could enable policymakers to anticipate which emerging capabilities pose the greatest risks.

- **Current assessment approaches are inadequate:** Existing approaches, such as vulnerability scoring systems like the Common Vulnerability Scoring System (CVSS) or threat intelligence, focus primarily on identifying which systems can be compromised, exploitability, and adversary behavior rather than downstream societal effects. The result is a fragmented landscape where technical capability assessments, strategic analysis, and policy evaluation operate in silos, limiting the ability for policymakers to anticipate and respond to high-consequence risks.
- **Implications analysis could inform policy triggers, mitigations, and evaluation priorities.** By connecting capability thresholds to likely societal outcomes, policymakers could make evidence-based decisions, define "red lines," design tailored and graduated responses, and direct investments toward those capabilities and attack pathways most likely to produce severe impacts.

## Suggested projects

The projects below are organized in three phases reflecting logical dependency rather than strict sequencing. Phase 1 (Foundations) builds an empirical evidence base. Phase 2 (Model Development) uses that evidence to construct capability-outcome models. Phase 3 (Application) translates those models into tools for policy design, safeguard prioritization, and evaluation targeting. Work across phases can proceed in parallel, but later phases benefit substantially from earlier outputs.

### Phase 1: Foundations

- **Capability bottleneck analysis:** This analysis identifies which gaps in offensive capabilities currently limit what attackers can do and examines how new capabilities, including AI-enabled tools, might remove or shift those limits. Capability bottleneck research can help clarify how different capabilities combine to enable new types of attacks, helping policymakers and stakeholders prioritize evaluations and defensive investments.
- **Historical operations mapping:** This analysis aims to trace causal chains from specific capabilities to observed societal effects across diverse operation types - from high-profile infrastructure attacks (Stuxnet, NotPetya, SolarWinds, Colonial Pipeline) to espionage and financially motivated campaigns. Mapping capabilities to impacts will help identify costs, mediating factors, and unintended consequences. Researchers may focus attention on what past cases reveal about how sudden capability developments affected the trajectory of offensive operations.
- **Structured operations dataset:** Researchers may wish to build a systematically coded repository of past cyber operations, recording standardized variables across cases, including: capabilities employed, target characteristics, attacker objectives and constraints, defensive posture, geopolitical context, and observed outcomes. Currently, knowledge of cyber operations is scattered across incident reports, academic case studies, and threat intelligence products with inconsistent framing, making systematic comparison difficult. A shared dataset would enable the cross-case analysis that Phase 2 modeling requires.

### Phase 2: Capability-Outcome Model Development

- **Capability-outcome pathway models:** Formal or semi-formal models could map pathways from capabilities to outcomes. Models should incorporate uncertainty, identify key dependencies, and highlight where small contextual changes produce large changes in outcomes. Researchers might consider drawing on risk propagation methods to represent cascade effects through interdependent systems.
- **Scenario-based projections:** Structured scenario exercises could explore how specific capability developments could translate into societal outcomes. Model development should engage diverse experts - technical, strategic, economic, and political - document reasoning chains, and incorporate sensitivity analysis to identify which assumptions drive outcome variance.
- **Indicators and warning frameworks:** Frameworks could help monitor signals that capabilities are being developed, proliferating, or being prepared for deployment. Such models can further provide early warning indicators of impending societal impacts.

### Phase 3: Application

- **Evaluation prioritization:** Frameworks could help determine which AI-cyber capability evaluations are most important and which are less critical by linking evaluation targets to their downstream societal significance rather than technical novelty alone.
- **Policy threshold development:** Capability-outcome models could guide the definition of concrete "red lines" or policy triggers tied to specific capability thresholds, enabling evidence-based responses to emerging capabilities rather than reactive policymaking.
- **Decision-support integration:** Decision-makers could pilot lightweight decision-support tools derived from the frameworks and integrate them within existing processes (threat assessments, vulnerability equities, policy reviews), while publishing case studies could document practical applications and methodological lessons for future use.

## Elements of success

These projects should aim not to perfectly predict societal outcomes - which remains infeasible - but to improve how analysts and policymakers reason about the downstream effects of specific capabilities. Even modest improvements in analytical rigor can strengthen evidence-based decision-making, reducing reliance on intuition or incomplete information. Successful projects will:

- **Build cumulative knowledge:** Document methodologies transparently, create shared datasets and case repositories, and design for extensibility so future work can build on prior efforts.
- **Engage policymakers and meet policy-relevant standards of rigor:** To inform real decisions, research outputs must meet the specific formats, evidence standards, and levels of rigor that policymakers require. This means working with (or under direction from) policy professionals from the outset, not translating academic work after the fact.
- **Differentiate by actor type:** Nation-states, criminal organizations, terrorists, and hacktivists have different objectives and constraints that produce different outcome distributions from identical capabilities. Frameworks should recognize that capability uplift could affect different actors differently and incorporate that into the analysis.

- **Account for adversary adaptation:** Static models become obsolete as defenders and attackers coevolve. Frameworks should model adversarial dynamics and be designed for regular updating.
- **Ground claims in evidence:** Analysis should distinguish theoretical capability from demonstrated operational effectiveness, acknowledge defensive countermeasures, and resist overstating threats. Credibility depends on maintaining this discipline throughout.

## Establish an AI security resource hub

Cost-effectiveness #1

### What is this?

A curated, open-source repository of AI security tools, threat intelligence, frameworks, and educational materials designed to help AI security practitioners focus their attention on trustworthy, effective security resources. The hub would consolidate existing efforts in one place and emphasize new and relevant elements with lightweight quality signals.

The hub would also include a limited builder on-ramp: a small, clearly labeled "Start here" section for software developers building with AI who need practical, secure-by-default guidance without becoming security experts. This on-ramp would focus on essentials - such as common failure modes, recommended baseline checks, and safe implementation patterns - and would explicitly point to deeper practitioner resources when needed.

### Why this matters

An AI security resource hub could help address an emerging gap between the pace of AI advancement and adoption and the maturity of AI security practices. AI capabilities that were unavailable until recently, including autonomous vulnerability discovery and exploitation, are arriving faster than most organizations are equipped to respond, making a collaborative effort to build better shared resources all the more urgent. This gap reflects several converging factors:

- **Organizations risk deploying AI without understanding security implications.** Organizations are integrating large language models and other AI systems into products and operations faster than security teams can evaluate new attack surfaces, increasing the chance of vulnerabilities being exploited.
- **Developers often lack clear, practical guidance on defenses.** Without easy access to frameworks, best practices, or evaluation tools, teams cannot reliably assess or implement security measures, leaving systems exposed.
- **Existing processes and standards are inconsistent or incomplete.** As noted in the introduction, the World Economic Forum's Global Cybersecurity Outlook[3] survey shows that while 66% of organizations expect AI to significantly impact cybersecurity in the year to come, only 37% report having processes in place to evaluate AI security, suggesting organizations might integrate AI systems into critical workflows without clearly understanding risks or available mitigations.

### Suggested projects

An AI security resource hub could provide practical, actionable support to help organizations understand, evaluate, and mitigate risks in AI systems. Potential projects to advance the development of a hub include:

- **Curated tool repository with lightweight validation cues:** Beyond listing tools, the hub would maintain basic quality signals such as maintainer notes, usage context ("best for X, not for Y"), last reviewed dates, and deprecation markers when tools become outdated. For a small subset of high-impact tools, deeper validation (benchmarks, independent reviews) could be added if resources allow.
- **Living literature reviews and links to incident and threat databases:** A continuously updated map of AI security research and real-world incident data, organized in parallel with the tool categories. Rather than creating new databases, the hub would aggregate and link to existing efforts such as MITRE ATLAS.
- **Training modules and workflow-oriented integrations:** If adoption and maintainer capacity permit, the hub could later expand to include practitioner-oriented training exercises (e.g., OWASP style labs) and workflow-oriented tooling integrations such as CI/CD checks or evaluation harnesses.

## Elements of success

To keep this feasible, the elements below are framed as a pragmatic progression from a lightweight initial version to a more robust version if adoption and resource availability permits. Successful efforts will:

- **Start small and stay useful:** The hub is most likely to succeed if it begins as a lightweight, practitioner-first resource that delivers immediate value. Early success can come from one or a small team of motivated curators, reserving more formal processes for if and when adoption demands it.
- **Include high-signal curation:** The core value is not comprehensiveness but selectivity - resources that practitioners actually use, with brief notes on what each is good for (and what it is not).
- **Prioritize findability over formal taxonomy:** Users should be able to locate relevant resources quickly through simple tags, categories, or collections. A consistent taxonomy can help, but it is not required; the goal is practical navigation and usability with minimal friction.
- **Include ways to highlight freshness and deprecation:** Each entry should show when it was last reviewed, with a straightforward mechanism to flag outdated items and deprecate resources that are no longer maintained or recommended.
- **Support a lightweight contribution workflow:** The hub should make contributions easy from the start through basic submission guidelines, an issue tracker, and maintainers who can accept or reject additions without heavy bureaucracy.
- **Use optional quality cues**: Simple labels such as "maintainer reviewed," "commonly used," or "benchmarked" can help developers quickly assess resources without implying that every resource has been formally audited.
- **Scale only when resourced:** Heavier efforts, such as deep validation of select tools, hands-on training modules, integrations, or additional languages, should be treated as optional upgrades that follow demonstrated demand and sufficient maintainer capacity.

## Clarify language through a standard taxonomy

Cost-effectiveness #9

### What is this?

Clear, consistent, and widely-adopted terminology, taxonomies, and conceptual frameworks could help build a stronger understanding of "AI security" and clarify its relationships to related concepts such as AI safety, trustworthiness, control, alignment, robustness, and reliability. Developing materials that support these frameworks could help researchers, policymakers, industry practitioners, and funders communicate more effectively.

### Why this matters

Precise, shared terminology could help accelerate learning and effective coordination. Although key terms - including "AI Security," "trustworthiness," "control," and "alignment" - are widely used, they often lack precise, consistent definitions and vary in interpretation across communities. For example, "AI Security" may refer to protecting AI systems from cyber threats - as in cybersecurity - or describing national security implications of advanced AI. "Alignment" can refer to alignment with a specific user or with broader societal values; "trustworthiness" may cover robustness, transparency, or a plethora of other AI ethics related issues; and "control" can refer to a very general problem of steering AI systems, or to a specific set of guardrails for artificial general intelligence.

Without common language, funders, researchers, practitioners, and policymakers risk operating in silos and missing opportunities to collaborate. Unclear terms could also degrade the quality of public discussions about AI.

Creating greater clarity would likely prove feasible: convening stakeholders to build even partial consensus could substantially improve coordination while requiring relatively modest resources.

### Suggested projects

Various efforts to clarify terminology exist across academic, industry, and policy contexts, but significant gaps remain. The following projects represent concrete opportunities to improve shared understanding:

- **Building a living glossary of AI terminology.** A Wikipedia-style resource could provide clear definitions, usage examples, common confusions, and cross-references for key terms. A glossary could acknowledge where multiple definitions coexist and explain the context for each. Unlike a static document, this resource should evolve as usage changes and new terms emerge.

- **Convening terminology workshops across research communities.** Workshops could bring together researchers from ML security, AI safety, alignment, trustworthy AI, and AI ethics to explicitly discuss terminology differences and work toward convergence where possible. These workshops should produce written outputs documenting areas of agreement and acknowledged areas of divergence.
- **Conducting large-scale terminology surveys.** Systematic surveys could gather insights from researchers, policymakers, and industry practitioners about how they currently use key terms, revealing patterns of usage across communities. This empirical data can ground consensus-building efforts and help identify the most important confusions to resolve.
- **Clarifying terminology in existing AI legislation and regulation.** Analyzing current and proposed AI statutes and policies (e.g., the EU AI Act, U.S. state-level AI bills, international frameworks) could help identify ambiguous terminology and prompt clarifications. This work could inform engagement with legislative staff to improve the precision and consistency of future policy language.
- **Developing accessible educational materials for policymakers, journalists, and communicators.** Guides and training resources could explain key distinctions in AI terminology without requiring technical expertise. These resources could use relevant examples and case studies for policy and media contexts, helping professionals accurately convey AI concepts (such as clarifying security vs. safety, alignment vs. capability, etc.) while avoiding sensationalism or understatement.
- **Testing comprehension of terminology clarifications.** Empirical evaluations could assess whether proposed taxonomies and definitions improve understanding by measuring comprehension across diverse audiences.

## Elements of success

Successful projects will:

- **Focus on high-impact confusion, not perfection:** Projects should limit the scope of terms and prioritize areas where miscommunication carries meaningful consequences, such as in regulation, research, or public discussion. Each effort should clearly state its scope and intended audience.
- **Acknowledge disagreements:** Some differences in terminology reflect substantive conceptual debates rather than simple misunderstandings. Projects should distinguish between disagreements that clearer definitions can resolve and those that reflect genuine differences in underlying assumptions or goals. In such cases, efforts should document and clarify those differences rather than attempting to eliminate them through terminology alone.
- **Prioritize adoption and upkeep:** Projects should not only define key terms but also plan for effective dissemination, collaboration, integration, and long-term maintenance in coordination with key stakeholders.
- **Avoid premature standardization:** Standardizing too early can suppress valuable diversity in approaches and thinking. Projects should aim to reduce harmful confusion while preserving space for legitimate debate in a rapidly changing field.

- **Build on existing work:** Efforts should review and incorporate prior initiatives from standards bodies, industry, and academia to avoid duplicating frameworks and increasing confusion.
- **Balance clarity and precision:** resources should remain accessible to their intended audiences while maintaining technical accuracy. Layered explanations - for example, providing one definition for a general audience and another for researchers - can help achieve this balance.

# Chapter 3: Public-private coordination and institutional infrastructure

Although AI labs face threats from sophisticated nation-state actors, they operate as private entities with limited access to classified threat intelligence, constrained resources for expensive security controls, and no established channels for coordinating responses to emerging risks. One pathway to address these gaps involves **enhancing private labs' coordination with governments**, which possess critical intelligence, can provide substantial security funding, and can facilitate information sharing across organizations. However, traditional coordination efforts, such as public-private partnerships, might not translate seamlessly to AI security: historically, these partnerships focused on contexts in which the government, as the primary developer of the capabilities, could embed security requirements from the start. Accordingly, traditional efforts might struggle to address the unique characteristics of AI security: a small number of private labs control critical capabilities, threat models evolve rapidly as AI advances, and global competition might discourage security investments perceived as competitive disadvantages.

**This chapter overviews priority areas that support effective cooperation between government and industry to secure AI systems.** It focuses on institutional arrangements that enable coordination while respecting the operational realities of both sectors: government agencies must protect sources and methods, while companies must maintain competitive positions and operational flexibility. These areas are as follows:

1. **Public-private partnerships for security investments** could allow governments to incentivize - and where necessary, mandate - private labs to adopt costly security controls that protect critical national assets.
2. **Enhanced background checks** could help labs assess personnel risks when hiring for positions with access to sensitive AI capabilities, particularly when facing sophisticated insider threats or foreign intelligence recruitment efforts.
3. **Information and incident sharing infrastructure** could establish channels through which the government can share threat intelligence with labs, and through which labs can report incidents and exchange defensive knowledge without competitive or legal concerns.
4. **AI incident response frameworks and playbooks** could help define protocols for coordinating across stakeholders during AI security incidents, from initial detection and containment to post-incident disclosure and remediation.

## Form public-private partnerships focused on AI security

Importance #6

### What is this?

Public-private partnerships (PPPs) for AI lab security represent collaborative frameworks between government entities and private AI companies to make large-scale investments in securing frontier AI systems against sophisticated threats. These partnerships address a fundamental challenge: the most advanced security measures required to protect transformative AI capabilities (particularly against nation-state adversaries) exceed what private companies would likely implement alone within reasonable timeframes.

In RAND's security framework taxonomy of Security Levels (SL)[16], SL-5 generally requires multi-year investment and government partnership. AI labs can reasonably achieve SL-3 (defending against moderately sophisticated attackers) within approximately one year and SL-4 (defending against highly sophisticated attackers) within 2-3 years through private investment. However, achieving SL-5 (defense against nation-state adversaries with billion-dollar budgets) requires an estimated minimum of five years and necessitates active support from national security communities.

Public-private partnerships in this domain take multiple forms:

- **Infrastructure partnerships** focus on developing secure computing facilities, potentially leveraging purpose-built secure datacenters designed to military specifications.
- **Knowledge-sharing partnerships** enable government cybersecurity expertise to red-team AI lab defenses, including threat vectors that are difficult for non-governmental actors to replicate.
- **R&D partnerships** support longer-horizon security research (e.g., advanced cryptography, secure systems engineering, and AI-specific defensive methods) that may be underprovided by private incentives.
- **Regulatory partnerships** establish security standards and compliance frameworks that incentivize or mandate appropriate security investments while ensuring these requirements remain technically feasible.
- **Financial partnerships** provide direct government funding or subsidies to offset the substantial costs of implementing advanced security measures.

### Why this matters

Ensuring SL-5 security for advanced AI systems requires measures far beyond what private firms can provide. The reasons government action is essential include:

- **Only the government possesses the unique authorities and capabilities required for SL-5 AI security.** SL-5 security requires air-gapped facilities at secure locations, counterintelligence support to detect and respond to state-sponsored infiltration attempts, and physical isolation measures that private companies cannot implement independently. The government uniquely possesses the authorities and capabilities that private firms cannot replicate at scale - such as designating and operating secure sites with classified protections, providing counterintelligence support, and establishing cross-organization regulatory frameworks for coordinated security.

- **Market forces do not adequately incentivize the investment necessary for national security-level protection.** Private firms face a structural disincentive to invest in SL-5 security: each additional layer of protection becomes much more expensive, while the benefits - such as reducing national-security risks - spread across society rather than the company itself. Market forces alone cannot account for these broader benefits. Government intervention (through co-funding, procurement guarantees, or shared infrastructure) can transform security from a cost to a strategic advantage.
- **Public-private partnerships provide a foundation for managing international AI security risks.** As AI security increasingly becomes an issue of international significance, public-private partnerships could prove critical in managing geopolitics on behalf of private labs, which cannot conduct foreign relations on their own. Any durable international approach to frontier AI security - whether focused on standards, verification, or incident response - will likely require government-to-government engagement that individual companies cannot conduct.

## Suggested projects

The following are illustrative project concepts that demonstrate the range of public-private partnership models in this space. Some have partial analogs in current programs, but they are presented here as candidate initiatives that stakeholders could launch or scale:

- **Secure computing infrastructure partnerships**: Private AI labs can enter into partnerships with the U.S. government that provide access to highly secure computing environments and expertise for protecting sensitive AI systems. The Department of Energy (DoE) and its national laboratories represent natural starting points for strategic government partnerships in secure AI development. DoE facilities already maintain high security standards and possess relevant expertise in protecting sensitive technologies, including Restricted Data, special nuclear materials, and other national security assets.[18] Early partnerships have focused on AI applications for scientific research, but the underlying infrastructure and security frameworks could scale to support frontier AI development with appropriate security measures.
- **Red team collaboration programs**: Private AI labs can partner with government cybersecurity agencies to test their defenses against sophisticated attack scenarios. These agencies possess offensive capabilities and threat intelligence unavailable to private sector security teams. Structured red team partnerships enable agencies to assess AI lab defenses against nation-state threats and provide feedback that helps labs identify and fix vulnerabilities before real adversaries can exploit them. These partnerships must protect sensitive government tools and intelligence, as well as the labs' infrastructure, all the while giving labs actionable guidance, but when properly structured, they could provide invaluable security validation.

- **Standards development initiatives**: Neutral third-parties, such as RAND, can convene government and industry stakeholders to develop security standards and frameworks. RAND's ongoing work defining SL-5 specifications involves a task force including AI labs, government security experts, and academic researchers. These collaborative standard-setting efforts could help ensure that security requirements remain technically feasible while still achieving necessary protection levels. They can also help create common frameworks that facilitate both regulation and industry best practices. Regular syncs between third-parties and AI labs on advanced R&D topics, such as secure datacenter exploration, illustrate how these partnerships could facilitate practical, ongoing collaboration.
- **Security-focused regulatory pilot programs**: Governments can run pilot programs that test security requirements and compliance mechanisms for frontier AI development and deployment. Examples include pilots for secure development environments (e.g., audited access controls for model weights and sensitive tooling), incident reporting and breach exercises, and evaluation/verification procedures tied to security levels. These pilots could help governments develop informed regulatory approaches while giving companies structured pathways to demonstrate compliance and harden operational security.

## Elements of success

A RAND analysis of U.S. government PPP models identifies several features directly applicable to AI lab security: shared objectives (e.g., maintaining the U.S. frontier AI lead and preventing theft of model weights), complementary capacities (where government provides secure infrastructure and counterintelligence while firms conduct R&D), cost- and risk-sharing, long-term contract horizons often exceeding twenty years, flexible agreement structures, structured incentives such as intellectual property protections and regulatory relief, and mutually determined governance.[19] Effective public-private partnerships for AI lab security will:

- **Establish a shared threat understanding**: Partnerships should begin with a shared understanding of specific threats and required security measures. While the RAND security level framework provides a useful starting point, partnerships should develop threat models specific to novel AI risks, deployment contexts, and adversary motivations. This requires ongoing intelligence sharing and joint analysis between government security agencies and private sector AI labs.
- **Align private incentives with public interest through cost sharing and incentive alignment**: Partnership structures should ensure companies do not bear disproportionate costs for security measures that primarily protect national security rather than their commercial interests. This might involve direct government funding for shared security infrastructure, tax incentives for security investments, or procurement commitments tied to verified security levels. Partnerships can also use non-financial approaches to align incentives, such as setting baseline regulatory security requirements across labs to create a level playing field, tying market access or deployment permissions to meet SL targets, and clarifying liability or safe-harbor rules for verified best-practice security.

- **Adopt governance and accountability structures:** Partnerships should require clear lines of authority and accountability. Governance frameworks should define decision rights over security standards, funding allocations, and operational oversight. Independent audit mechanisms (potentially through a joint government-industry board) can ensure transparency and prevent either side from dominating strategic decisions. Without such structures, partnerships risk becoming ad hoc or politically vulnerable.
- **Plan for long-term commitment**: The five-plus-year timeline for achieving SL-5 security demands sustained commitment from both government and industry partners. Political leadership changes, shifting budget priorities, and market pressures all threaten continuity. Partnerships should establish institutional structures with bipartisan support, multi-year funding commitments, and governance frameworks that can persist across administration changes and market cycles.
- **Include international coordination mechanisms**: Given the global nature of both AI development and security threats, partnerships could benefit from including mechanisms for international coordination. This might involve multilateral working groups on AI security standards, intelligence sharing agreements between allied nations regarding AI-related threats, or international treaties establishing security requirements and verification mechanisms. The challenge lies in balancing the benefits of international transparency and coordination against legitimate competitive and national security concerns.
- **Embed AI security across innovation ecosystems:** Success will depend on embedding security partnerships into the national innovation ecosystem. Partnerships should link together AI research funding, defense modernization, and industrial policy. Delays in establishing these partnerships could create not only technological risks but also strategic vulnerabilities for national security.

## Adapt enhanced government background checks for sensitive AI lab personnel

Cost-effectiveness #10

### What is this?

Government personnel security practices - long used in defense and intelligence - can be adapted for private AI labs developing frontier AI systems. Formal programs - including Facility Clearances (FCL) and Personnel Security Clearances (PCL) - adapted for private AI labs could enable them to vet employees and contractors and to support ongoing monitoring and structured intelligence cooperation for those handling sensitive model weights, training data, proprietary codebases, novel algorithmic techniques, and research with national security implications.

These measures go substantially beyond standard commercial background checks, which typically cover only criminal records and employment verification. Legal authorities allow deeper inquiry and impose stronger obligations on participants. Government-supported security clearance programs conduct comprehensive investigations into an individual's financial stability, foreign contacts and travel, substance use history, mental health treatment patterns, employment history, and social media activities.

The process requires extensive documentation through government forms, fingerprinting, and potential polygraph examinations. In some cases, false statements can carry legal consequences beyond employment-related repercussions. While not all elements transfer to non-classified work in the private sector, more extensive background checks could help reduce insider threats.

This framework should include multiple security levels with clearance designations determining access to sensitive channels, systems, and model weights.

### Why this matters

AI labs increasingly face threats that go beyond typical commercial risks, resembling those of defense contractors and critical infrastructure operators:

- **Nation-state actors are actively seeking access to U.S. AI technologies through insider recruitment and deceptive remote-hiring schemes.** For example, North Korean operatives have obtained U.S. IT positions using stolen identities and AI-generated images, sometimes running “laptop farms” to gain access to company systems.[20] Similarly, a former Google engineer was convicted of stealing thousands of pages of AI trade secrets for the benefit of China, including custom hardware designs and software enabling advanced AI training.[21] These incidents highlight that conventional background checks and identity verification processes may be insufficient to prevent infiltration by highly capable, state-aligned actors, underscoring the need for enhanced vetting and clearance mechanisms for personnel with access to sensitive AI systems.

- **Insider threats are among the most significant risks for AI model weight theft and intellectual property compromise:** The 2024 RAND "Securing AI Model Weights" report explicitly identifies reducing the number of authorized personnel with access to sensitive systems as a priority security measure within their Security Level framework. The stakes remain high: model weights carry both immediate commercial value (potentially worth billions) and strategic national security implications. A single compromised insider - whether compromised through financial pressure, foreign recruitment, or ideological motivations - could exfiltrate years of research in minutes.
- **Current commercial background check systems used by most technology companies are insufficient to address these threats**. These checks cannot identify financial vulnerability to blackmail, problematic foreign intelligence contacts, or susceptibility to coercion. The holistic approach used in government security clearances examines the "whole person," assessing whether an individual is trustworthy, reliable, and not vulnerable to foreign intelligence exploitation.
- **The urgency to protect sensitive AI assets grows as labs must increasingly handle classified information when partnering with government agencies on national security projects.** Without appropriate vetting infrastructure, companies cannot bid on classified contracts, compartmentalize sensitive work, or access classified threat intelligence that could strengthen their security. Expanding clearances or "clearance-like" mechanisms for AI companies could improve compartmentalization and defensive capabilities. One approach would extend national security clearance processes to AI researchers in sensitive roles; another would conduct tailored background investigations using select components of the clearance process (such as foreign-intelligence association checks) without issuing a formal clearance.[22]

## Suggested projects

Several concrete initiatives illustrate potential project areas through which governmental support for personnel security in AI labs could advance. Some of these efforts are nascent or partially underway, while others would require new policy, legal, or institutional work to fully realize.

- **Legislative framework development:** Proposals such as Section 1627 of the FY2026 National Defense Authorization Act[23] on physical and cybersecurity procurement requirements for AI systems illustrate how legislation could establish risk-based security frameworks addressing insider threat, supply chain, and workforce protection requirements for entities contracting with the Department of Defense on AI technologies. Advancing this work would involve resolving open questions around presidential designation criteria, enforcement mechanisms, trade secret protection, and how companies share information with government agencies.
- **AI Security Center collaboration:** Expanding the role of the AI Security Center (AISC), in coordination with CISA and the intelligence community, could provide AI companies with actionable insider threat intelligence. This could include mechanisms for surfacing unclassified information on nation-state tactics, techniques, and procedures (TTPs) targeting AI laboratory personnel.

- **Facility clearance programs:** Supporting AI companies in establishing and scaling Facility Clearance (FCL) capabilities through DCSA could involve developing operational infrastructure - policies, training curricula, clearance processing workflows, and compliance systems needed to sustain FCLs over time. It could also include building a professional ecosystem of Facility Security Officers (FSOs) and clearance support services.
- **Internal clearance systems:** Designing internal access control and clearance-like systems for AI labs outside federal classified programs could help protect labs against insider threats. Stakeholders could explore legal authorities to extend some elements of government clearance regimes into private employment. This work would define best practices for tiered access to sensitive systems, model weights, and proprietary information.
- **Advanced security level standards:** a standards-development project could advance and formalize higher-tier security frameworks such as SL-5, incorporating comprehensive personnel security requirements as part of a defense-in-depth model. This would build on ongoing efforts led by organizations such as the Institute for Science and Technology.

## Elements of success

Effective governmental support for personnel security in AI labs requires several critical components working in concert. Successful projects will:

- **Ensure sustained public-private collaboration**: Sustained collaboration between federal agencies (DCSA, FBI Counterintelligence Division, CISA, NSA AI Security Center) and private companies should include formal clearance sponsorship, ongoing threat intelligence sharing, and security incident coordination. It should also support the joint development of industry-appropriate security standards that protect sensitive assets while allowing AI labs to continue to innovate.
- **Establish dedicated security personnel and expertise**: AI companies should invest in Facility Security Officers with expertise in personnel security, clearance adjudication, and compliance requirements. Many companies will need external partnerships with specialized personnel security firms to provide adequate coverage, expertise, and continuity during transitions or absences. Legal frameworks should support extending these programs into non-classified contexts.
- **Apply holistic vetting approaches**: Personnel security programs should examine multiple risk dimensions: financial stability (vulnerability to bribery), foreign contacts and preferences (foreign intelligence risk), criminal history, substance use patterns, employment consistency, and social media behavior. Such holistic assessments focus on trustworthiness and vulnerability to coercion.
- **Conduct continuous monitoring and periodic re-vetting**: Rather than relying solely on initial background checks, companies should update them on a periodic basis or when there is a significant change in an employee's access to or control over sensitive information. This will help ensure ongoing suitability and detect changes in financial status, foreign travel patterns, or behavior that could indicate compromise.
- **Rigorously apply the "need-to-know" principle**: Individuals' clearance levels should match or exceed the sensitivity of information they access. This requires careful information classification, role-based access controls, and technical enforcement mechanisms to prevent unauthorized access even by cleared personnel to information beyond their role requirements.

- **Clarify legal and regulatory frameworks**: Governments should provide industry statutory authority, standardized criteria for which AI firms must comply with personnel security requirements, transparent enforcement mechanisms, and protection for proprietary information shared with government agencies during security processes. Clear frameworks can help reduce uncertainty around presidential designation authority and compliance consequences.
- **Foster a culture of following security best practices**: In addition to relying on personnel compliance, organizations should build a security culture where employees understand threat models, recognize social engineering attempts, report concerning contacts proactively, and view security measures as mission-enabling. This may require commitment by senior organizational leadership, regular training, and integration of security considerations into operational decision-making.
- **Balance scalability with operational needs:** Investigation and clearance practices should protect sensitive systems while enabling collaboration, knowledge sharing, and research progress. Overly restrictive personnel security can create single points of failure, slow research, and make organizations less attractive to top talent.

## Advance information and incident sharing

Importance #2, Cost-effectiveness #6

### What is this?

Information-sharing - and the organizational, legal, and technical infrastructure needed to support it - offers a potential pathway to advance AI security. Governments and AI labs could adapt proven information-sharing practices used in other high-risk sectors, including dedicated coordination organizations, governance frameworks, and supporting technical and operational processes. These practices would enable AI developers, major deployers, and government stakeholders to share information about incidents, near misses, vulnerabilities, emerging threats, and defensive lessons.

This work considers a range of established coordination mechanisms, such as an AI-specific Information Sharing and Analysis Center (ISAC), government-facilitated forums, industry-led sharing consortia, or hybrid public-private arrangements.

The scope extends beyond narrow threat intelligence. It includes AI-specific failure modes, attack patterns, operational lessons, and incident-response practices that are currently learned in isolation and shared inconsistently, if at all.

### Why this matters

Effective information and incident sharing are critical to defend against fast-evolving threats in the frontier AI ecosystem. The reasons new, purpose-built information and incident sharing mechanisms are needed include:

- **Information sharing enables organizations to collectively address shared threats that cannot be managed in isolation.** In the frontier AI ecosystem, organizations face shared threats including nation-state intrusion, supply-chain compromise, insider risk, data poisoning, model theft (including distillation), and misuse of deployed systems. Rather than respond to these threats independently, information-sharing could enable these organizations to address them collectively.
- **Existing channels are not designed to capture and disseminate AI-specific incidents across all relevant organizations.** Although general cybersecurity information-sharing mechanisms and government channels exist, they are not designed to consistently capture, redact sensitive details from, and disseminate AI-specific incidents, vulnerabilities, and operational lessons across competing frontier labs and deployers. As a result, failures or attacks discovered by one organization often remain unknown to others until they recur.
- **The unique concentration of expertise and rapid technological change in frontier AI make fast, specialized sharing essential.** Unlike other traditional sectors, only a small number of frontier AI labs possess much of the most security-critical knowledge, yet few established norms or channels govern how they share with each other or with downstream deployers. At the same time, the rapid evolution of technology itself compresses the window for useful information exchange: vulnerabilities discovered today may become irrelevant within months as models and deployment patterns shift, increasing the importance of faster information-sharing relative to other domains. Effective analysis might further require expertise spanning machine learning and cybersecurity - a skillset that remains limited and unevenly distributed. Together, these factors mean that existing, fragmented sharing practices are not just inefficient but structurally mismatched to the speed and specialization the threat environment demands.

- **There is a lack of institutions and technical infrastructure to operationalize timely, protected information sharing.** The primary challenge is not awareness of the problem, but the lack of institutions and technical systems to address it. Partial organizational infrastructure exists, such as the Frontier Model Forum, which facilitates some coordination among a subset of leading AI companies. But gaps remain in participation, security-specific depth, and structured operational sharing. Without durable mechanisms for protected and timely information exchange that complement and build on these early efforts, the AI ecosystem cannot keep pace with the speed or impact of frontier model development.

## Suggested projects

A path to closing the information-sharing gap in frontier AI is the establishment of a dedicated coordination organization, such as an AI-focused Information Sharing and Analysis Center (ISAC) or equivalent body, that can integrate the functions described in this priority area under a single professional institution. ISACs and similar organizations have proven effective in finance, energy, healthcare, and other critical sectors precisely because they combine governance, reporting infrastructure, operational coordination, and government interfaces in a durable, trusted structure. A dedicated AI security sharing organization could be far more effective than any of these functions pursued in isolation.

In the absence of a central information-sharing body, additional individual projects can enable enhanced information-sharing or help foster an environment in which such a body might be more easily established. Potential projects include:

- **AI security incident taxonomy and gap analysis.** Clearly defined and refined taxonomies of AI security incidents can help clarify which scenarios apply to providers, deployers, or targets. Analysis that maps these against existing incident response and reporting protocols can identify where current frameworks fall short, whether due to missing detection methods, unclear escalation paths, or undefined remediation steps. This foundational work is a prerequisite for nearly all other efforts in this space.
- **Privacy-preserving incident sharing mechanisms.** Anonymization, aggregation, and access control methods could enable organizations to share incident details with trusted peers or coordination bodies without exposing proprietary model information or sensitive operational data.
- **AI incident dataset curation.** Assembled and curated datasets of real-world AI incidents could support empirical research, tool development, and playbook validation. Datasets should capture a range of incident types, actor roles, and outcomes, and should be structured to enable comparative analysis across the taxonomy developed above.

- **Governance and legal enablement.** Stakeholders could explore governance models for AI-focused information-sharing mechanisms, including membership criteria, tiered access, funding structures such as membership fees, and decision-making processes. Research should address legal barriers such as antitrust concerns, liability exposure, and contractual constraints by developing participation rules, reporting obligations, or safe-harbor frameworks that support good-faith security and safety sharing.
- **Incident and near-miss reporting frameworks.** Stakeholders could develop standardized approaches for describing and classifying AI-relevant incidents and near misses, with clear handling, protection, and escalation rules. Designs should balance timely, actionable insight with safeguards for sensitive or proprietary information, drawing on practices from other high-reliability sectors while adapting to AI-specific risks. This work is closely linked to the AI Incident Response priority area that closes out this chapter, which addresses how organizations detect, manage, and learn from incidents once they occur.
- **Coordinated response and information fusion.** Stakeholders could design options for shared operational capabilities, such as joint incident-response playbooks, analyst coordination forums, or fusion-style processes that can operate within an ISAC or similar coordination body when incidents affect multiple organizations.
- **Interfaces with government partners.** Stakeholders could design structured, repeatable interfaces for two-way exchange with relevant public-sector actors, building on existing coordination mechanisms such as CISA's Joint Cyber Defense Collaborative.[24] This includes mechanisms for sanitizing, aggregating, or tiering sensitive information and for acting on time-sensitive alerts, without requiring direct disclosure of proprietary model details.
- **AI-specific extensions to existing security frameworks.** Stakeholders could work with established standards bodies and information security frameworks - such as MITRE ATT&CK and STIX/TAXII[25] - to better capture threats and failure modes unique to AI systems, including attack patterns against model weights, training pipelines, and inference infrastructure. The priority should be on extensions and mappings rather than bespoke or parallel taxonomies.
- **Participation incentives and sustainability.** Stakeholders could design sustainability models that reflect standard practice in other sectors, including membership fees, contribution expectations, and, where appropriate, incident-reporting requirements tied to participation benefits.

## Elements of success

To meaningfully reduce risks in the frontier AI ecosystem, information and incident sharing should move beyond high-level promises of cooperation and establish lasting, practical systems for coordination, legal clarity, and day-to-day operations. Successful projects will:

- **Advance institutional credibility and trusted participation.** Effective sharing will require institutional credibility - including clear governance, transparent decision-making, and demonstrated value to participants - to sustain participation from frontier AI developers, major deployers, and relevant public-sector partners. Participants should consider strategies to advance trust, such as through predictable handling of sensitive disclosures, consistent application of rules, and confidence that participation will not create disproportionate legal, reputational, or competitive risk.

- **Establish clear scope and boundaries for shared information.** Sharing infrastructure should define what types of information are in scope - such as incidents, near misses, vulnerabilities, and AI-specific failure modes - and what is explicitly out of scope. These boundaries can help protect proprietary information, separate security and safety collaboration from product or commercial coordination and prevent mission creep. Scope definitions should be precise enough to guide reporting and use, while flexible enough to evolve as AI risks change.
- **Establish strong protections for sensitive information.** Sharing systems should embed strong protections for sensitive information through enforceable handling rules, access tiers, and use restrictions. Technical and organizational controls should ensure that shared information reaches only appropriate participants and is used solely to reduce risks. Legal safeguards, such as liability protections for good-faith reporting, could help encourage participation. The sharing infrastructure itself should be treated as a high-value asset, with its technical systems, data stores, and access pathways strongly secured, since any compromise or misuse could undermine trust and increase, rather than reduce risk.
- **Produce and disseminate actionable, timely, and operationally relevant outputs.** Information sharing should prioritize speed and usefulness over completeness. Effective systems emphasize early warning, practical indicators, and mitigation guidance that participants can act on during ongoing incidents or emerging threats. Processes should support rapid dissemination and coordinated response when risks affect multiple organizations, rather than functioning solely as retrospective reporting channels.
- **Integrate with existing security and governance practices.** To be sustainable, AI information-sharing infrastructure should integrate with participants' existing security operations, incident-response processes, and governance frameworks. Alignment with established cybersecurity practices and standards reduces duplication and increases the likelihood that shared information is incorporated into real operational decisions rather than treated as external or advisory.
- **Sustain long-term participation and operations.** Durability requires stable funding, clear contribution expectations, and mechanisms to prevent free-riding. Standard approaches from other sectors, such as membership fees, contribution norms, or reporting requirements tied to access benefits, help ensure that the sharing infrastructure can operate continuously and evolve over time as risks and participation expand.

## Develop incident response frameworks and playbooks

Cost-effectiveness #3

### What is this?

Organizations could develop **operational frameworks and playbooks to guide responses to security incidents** involving AI, whether AI serves as the target, the tool used to perpetrate an attack, or a component of the affected infrastructure. Mechanisms that coordinate information and incident sharing can integrate these frameworks and playbooks to help organizations prepare for, detect, contain, remediate, and learn from incidents that involve models, data pipelines, or AI-enabled applications.

Work in this area could adapt existing incident response best practices to AI-specific risks and clarify how different actors should respond depending on their role in the AI supply chain. For example:

- Organizations deploying or operating AI systems may need to contain compromised integrations, disable affected applications, or coordinate with providers.
- AI providers may need to remediate or retrain models, isolate inference endpoints, revoke API access, freeze fine-tuning pipelines, roll back to known-safe model checkpoints, or restrict agent capabilities.
- Organizations targeted by AI-enabled attacks may require guidance on detection, attribution, and coordination with law enforcement or regulators.

Frameworks and playbooks can further clarify when and how AI incidents should be reported or shared with regulators, partners, and trusted peers, and how to distinguish between incidents that warrant internal handling and those requiring coordinated disclosure.

### Why this matters

Existing incident response models are not tailored to the unique risks of frontier AI. AI systems increasingly function as critical infrastructure, and incidents can have cascading effects across organizations and sectors. Yet current response practices are fragmented and sometimes improvised. While some incidents, such as data leakage or privacy breaches, already align with established practices, other AI-specific types of incidents - model integrity attacks, model theft, adversarial input manipulation, and misalignment incidents - fall largely outside existing frameworks and lack clear detection, attribution, or remediation pathways.

This uneven coverage highlights the need for dedicated incident response research and standardized playbooks that address AI-specific risks. Coordinated frameworks can help ensure that different types of actors know when and how to escalate, contain, and communicate during an AI-related event.

## Suggested projects

Numerous efforts already exist to improve AI incident response, each addressing different aspects of the problem. Given this fragmented landscape, it is important to review the current state of the field before launching any new initiative. One impactful output would include a set of comprehensive, operationally validated AI incident response playbooks adopted across the ecosystem, tailored to different roles in the AI supply chain. Building toward improved AI incident responses requires foundational research, empirical testing, and standardization. Individual projects that can deliver additional value toward this broader effort include:

- **AI-focused incident response exercises.** Tabletop and operational exercises that simulate AI-specific scenarios, such as model theft, adversarial input attacks, or compromised training pipelines, can test detection and coordination mechanisms across provider and deployer roles. Exercises should surface gaps in existing playbooks, clarify decision authority, and build operational familiarity among responders.
- **AI-specific detection, investigation, and response tooling.** Stakeholders could develop tools that support AI-specific stages of the incident response lifecycle, including detection of model integrity violations, investigation of training data compromise, and automated containment actions such as endpoint isolation or rollback to known-safe checkpoints. Tools should be designed for use across both provider and deployer contexts.
- **Standardized AI incident response playbooks.** Stakeholders could codify incident response procedures tailored to different roles, including providers, deployers, and regulators. Playbooks should address AI-specific decisions such as when to freeze fine-tuning, revoke API access, or roll back model versions, and should be validated through empirical studies and operational exercises before broad adoption.

## Elements of success

The elements below describe interdependent conditions for effective AI incident response and reporting. They can be developed incrementally but deliver the greatest value when implemented together and reinforced through operational use. Effective efforts will:

- **Clarify roles across the AI supply chain.** Frameworks should specify how model developers, platform providers, and downstream deployers coordinate during incidents, including clear handoffs and escalation paths.
- **Provide actionable decision authority.** Effective response depends on clearly defined authority to take disruptive actions, whether isolating endpoints, pausing fine-tuning, or disabling affected applications.
- **Integrate with organizational risk governance.** Incident response should connect directly to existing risk management and compliance structures, aligning with frameworks such as the NIST AI RMF and ISO/IEC 42001.
- **Leverage trusted mechanisms for learning and sharing.** Long-term risk reduction depends on structured sharing mechanisms that enable timely coordination, protect sensitive information, and complement regulatory disclosure requirements.

# Chapter 4: Technical security engineering and assurance

AI systems face increasingly sophisticated threats, and protecting critical assets - such as model weights and novel AI infrastructure - will require solutions beyond traditional software approaches. **This chapter addresses the technical infrastructure, methodologies, and specifications required to secure AI systems in practice.** While the preceding chapters establish strategic frameworks and institutional coordination mechanisms, meaningful AI security ultimately depends on building systems that can withstand sophisticated attacks and developing rigorous methods to verify their effectiveness. Current AI deployments often rely on security approaches borrowed from traditional software systems - approaches that may prove insufficient for protecting assets such as model weights, which represent billions of dollars in training investment and could enable significant harm if stolen or tampered with. This chapter emphasizes practical engineering and rigorous verification: what can actually be built with current technology or new capabilities, how to assess whether it works, and what specifications ensure that security investments produce genuine protection rather than security theater.

The following priority areas include high-level technical approaches for strengthening AI security across the system life cycle, from runtime protection to code generation, assessment, and adversarial testing:

1. **Confidential computing solutions** protect AI systems during runtime, ensuring that weights remain secure during inference, even against attackers with access to the underlying infrastructure.
2. **Secure code generation** explores whether AI itself can help write more secure code, potentially addressing the software vulnerabilities that currently enable many attacks against AI infrastructure.
3. **AI security posture assessment** provides frameworks for systematically evaluating the extent to which organizations protect their AI assets.
4. **Secure weight modules** examine dedicated hardware solutions for protecting model weights - arguably the most valuable and hardest-to-replace component of frontier AI systems.
5. **Elite security red-teaming specifications** define rigorous adversarial testing standards capable of identifying critical vulnerabilities before adversaries exploit them.

## Implement confidential computing to protect key AI system components

Importance #4

### What is this?

As critical components of AI systems, such as model weights and training code, increasingly come under threat of theft or unauthorized access, private labs and governments could turn to stronger technical safeguards. One solution includes **confidential computing,** a set of security practices that protect sensitive data even while a system processes it. This approach runs workloads inside hardware-based trusted execution environments (TEEs) - secure, isolated areas within a computer's processor that shield data and code from the rest of the system. Unlike traditional security approaches that protect data while it sits in storage (through encryption) or when it moves across networks (via Transport Layer Security - TLS), confidential computing adds another layer of protection while data is actively in use, reducing the risk that attackers can view or tamper with sensitive components during computation.

In the AI context, confidential computing typically involves several key technical elements, including hardware-rooted attestation mechanisms that prove specific code is running in a secure environment; memory encryption that prevents unauthorized access to sensitive data during processing; and constrained input/output channels that prevent accidental or malicious data exfiltration. Major cloud providers and chip manufacturers now offer TEE capabilities through technologies such as AWS Nitro Enclaves, Intel SGX and TDX, AMD SEV-SNP, and other accelerator technologies for GPUs and NPUs.

Confidential computing for AI involves more than merely running a model inside a secure enclave during inference. Production-ready approaches must secure the entire lifecycle: encrypting prompts before they enter the system, protecting model weights loaded into accelerator memory, isolating inference computation from host systems, and ensuring outputs remain encrypted until delivered to authorized recipients. Post-training processes - fine-tuning, alignment adjustments - require equivalent protection.

This requires careful architectural design to balance security guarantees with the operational realities of large-scale AI systems, including considerations for model loading, inter-node communication in distributed inference, monitoring and observability, and performance optimization.

### Why this matters

Confidential computing has emerged as a critical priority for AI security based on converging technical, regulatory, and competitive pressures. Leading AI labs have independently identified confidential computing as essential for protecting their most capable models: OpenAI has called for extending trusted computing to AI accelerators,[26] Anthropic published a joint whitepaper[27] on confidential inference systems, and Google DeepMind's Frontier Safety Framework[28] specifies confidential computing capabilities for highest-risk models. This convergence reflects recognition that traditional perimeter-based security is insufficient when model weights must be decrypted during use, creating vulnerability windows that only hardware-based trusted execution environments can close.

The need for confidential computing stems from concrete security challenges:

- **Advanced AI systems increasingly require protection mechanisms commensurate with their capability levels**. Frontier AI safety policies[29] explicitly require enhanced security controls - possibly including confidential computing - for models assessed at ASL-4 and above. These controls aim to prevent model weight exfiltration by sophisticated threat actors, insider threats, or supply chain compromises that could enable unauthorized replication of highly capable systems. Without hardware-based isolation, security teams face significant challenges in preventing accidental disclosure through insiders and errors.
- Enterprise and government customers increasingly demand stronger security guarantees before deploying AI systems for sensitive workloads. Organizations in regulated industries often require assurance that their data remains inaccessible to the AI provider's personnel and infrastructure. Confidential computing provides a technical foundation for these assurances through cryptographic attestation: customers can verify that their data enters secure enclaves and can only be processed by attested code, with no privileged access even for cloud provider administrators. This capability is becoming a competitive differentiator and, in some early cases, a prerequisite for certain customer segments.
- **Modern AI deployment practices create structural vulnerabilities that confidential computing can mitigate.** When AI companies partner with cloud providers to offer managed services, sensitive information flows across organizational boundaries. This includes not only customer data but also proprietary operational know-how about running large language models at scale. Confidential computing enables partnership models where providers can deliver high-quality AI services without requiring extensive access to partner infrastructure or intellectual property, reducing concerns about competitive knowledge transfer. Beyond cloud-hosted inference, confidential computing opens additional deployment modalities: on-premises inference on customer servers for enterprises with data residency requirements, and edge computing on ISP infrastructure for latency-sensitive applications. In each case, closed-weight providers can execute on hardware they do not own while retaining control over proprietary weights.
- **Confidential computing serves as a "race-to-the-top" coordination point.** Unlike many security measures that create competitive disadvantages, it allows organizations to credibly commit to strong security through verifiable technical controls, creating opportunities for industry-wide improvements as hardware support matures.

## Suggested projects

Practitioners and researchers can advance confidential computing for AI through several high-impact project areas:

- **Lab-cloud provider partnerships for confidential inference systems:** Major collaborations are underway between AI labs and cloud providers to bring confidential computing to production inference: Microsoft and OpenAI on Azure, Amazon and Anthropic on AWS. Practitioners could join these efforts, contribute independently to the open-source components of these systems, develop interoperability layers between different TEE implementations, or build tooling that addresses shared challenges such as encryption proxies, key management at scale, or performance profiling for confidential GPU workloads.
- **Standardized attestation frameworks:** The ecosystem currently lacks common attestation standards for verifying that different TEEs - such as AMD SEV-SNP and NVIDIA H100 confidential mode - are secure and functioning as intended. Projects could develop unified attestation formats and verification tooling (with an emphasis on relevant AI technologies such as H100s), create reference implementations for AI-specific key management systems, or build observability tools that provide necessary monitoring capabilities while respecting confidential computing boundaries. Success here would significantly reduce the engineering burden for organizations implementing confidential AI and accelerate industry-wide adoption.
- **Benchmarks and performance characterization:** Production deployment decisions require rigorous performance data across different hardware configurations, model architectures, and workload patterns. Projects could develop comprehensive benchmarking frameworks for confidential AI workloads, characterize the performance impact of different security configurations, identify optimization opportunities in the software stack, or establish baseline metrics that help organizations make informed tradeoffs between security guarantees and operational requirements.
- **Integration of confidential computing into training and research workflows:** Current confidential computing efforts focus primarily on inference, but training pipelines handle equally sensitive assets: proprietary datasets, novel architecture, alignment techniques, and intermediate checkpoints representing months of compute. Projects could develop frameworks for confidential fine-tuning or design secure pre-training architectures. This is particularly relevant for frontier labs under responsible scaling commitments, where protecting the research process - not just deployed models - becomes essential as capability levels increase.

## Elements of success

Projects advancing confidential computing for AI are most likely to succeed when they:

- **Identify and solve barriers to real deployment:** A strong measure of success lies in whether the work results in deployed security controls that protect real systems and users. The gap between confidential computing demos and production systems is substantial. Projects should identify specific barriers that prevent organizations from deploying confidential AI workloads today, whether complexity of key management at scale, difficulty integrating attestation into existing CI/CD pipelines, lack of monitoring tools that work within confidentiality boundaries, or uncertainty about how to handle incident response without inspecting enclave state. Projects that remove concrete blockers for real deployments create more value than those advancing capabilities that are not yet bottlenecks.

- **Work within hardware realities:** Effective projects target existing capabilities of available TEE hardware (NVIDIA H100, AMD MI300 with SEV-SNP, AWS Nitro and Trainium) rather than features that could emerge in the future. Stakeholders implementing confidential computing should understand each platform's security properties, identify gaps, and distinguish what is achievable today from what depends on next-generation hardware. Projects that help characterize real-world hardware behavior - documenting security boundaries, identifying edge cases, and developing practical workarounds - provide immediate value and could help accelerate broader deployment.
- **Build shared infrastructure:** Organizations often share many confidential computing challenges, including attestation verification, key management, secure logging, compliance documentation. Projects that contribute to reusable tooling, common standards, or reference implementations - such as cross-platform attestation formats or verification tooling for auditors - could help the entire ecosystem adopt confidential computing more quickly.

# Research secure code generation to write more secure code

Importance #5

## What is this?

Secure code generation research focuses on **developing and advancing AI techniques that improve the security of AI-produced code**, either by strengthening existing approaches or by creating new methods that help prevent vulnerabilities and insecure practices.

## Why this matters

The increasing prevalence of AI-generated code is transforming software development, creating both opportunities and risks for security. The following points illustrate why research on secure code generation is critical:

- **AI-generated code now constitutes a significant portion of software development.** By early 2026, about 50% of new code at Google was AI-generated.[30] GitHub Copilot surpassed 20 million users by mid-2025, with 90% of Fortune 100 companies adopting the tool.[31] Y Combinator noted in early 2025 that a quarter of their Winter 2025 batch startups had codebases that were at least 95% AI-generated.[32] The emergence of agentic coding systems such as Claude Code, Cursor, and GitHub Copilot Workspace is accelerating this trend.
- **AI performance on real-world engineering tasks has improved dramatically.** In early 2024, AI systems solved fewer than 14% of real-world coding tasks correctly. By late 2025, top models correctly solved more than 75% of tasks, showing dramatic gains in reliability and making AI-generated code a larger part of real software development.[33]
- **AI-generated code creates both opportunities and risks.** AI tools can accelerate vulnerability patching, provide real-time security guidance, and encourage safer programming practices such as memory-safe languages. However, empirical evaluations show that current AI systems often produce insecure code,[34] and if experienced developers rely too heavily on AI while newer programmers generate code without adequate security knowledge, vulnerabilities could spread widely across projects.
- **High-value targets increasingly depend on AI-generated code.** Major enterprises, government agencies, and other actors are adopting AI coding tools, meaning insecure AI-generated code could affect high-value targets at scale.
- **The scope of risk extends beyond application code.** Infrastructure-as-code (Terraform, CloudFormation, Kubernetes manifests), CI/CD pipeline configurations, API specifications, and security policies are increasingly AI-generated, which could introduce misconfigurations, overly permissive access controls, and exposed secrets.
- **Existing tools could produce rapid, actionable results** Established benchmarks can help measure results quickly, allowing small teams to advance security techniques that can be incrementally adopted across the industry.

## Suggested projects

This space is moving fast, making it especially important to first survey the landscape for existing efforts, including Meta's CyberSecEval benchmarks[34], academic security evaluation frameworks, and commercial tools addressing aspects of this space. Additional projects to complement improvements in ongoing efforts include:

- **Public benchmark and leaderboard:** A standardized benchmark suite for secure code generation with regularly updated evaluations could help measure both functional correctness and security properties. These tools could create incentives for improvement and enable rigorous comparison across systems.
- **AI security code generation capture-the-flag (CTF) competitions:** CTF competitions could help promote secure code generation. Teams could alternate between designing services (Blue Team) and crafting attacks (Red Team), providing adversarial testing that accelerates discovery of real-world vulnerabilities and defense strategies. The resulting data could be used for further research into post-training techniques to improve the quality of code safety generated by models.
- **Post-training security enhancement research:** Techniques to improve security of AI-generated code after initial training could include a broad range of methods, such as Reinforcement Learning with Verification Feedback (RLVR), which integrates automated verification and program analysis tools into the training loop to optimize robustness.
- **Test-time compute for security:** Research has established that additional inference-time compute improves code quality, but it has yet to establish how this applies specifically to security. Additional research could investigate which inference strategies (chain-of-thought, search, self-critique) yield the largest security improvements, how gains scale with computing budget, and what combinations of prompting strategies and verification tools are most effective.
- **Automated codebase migration to memory-safe languages:** AI systems could advance the translation of legacy C/C++ codebases into memory-safe languages like Rust. These efforts can coordinate with, or build upon, major initiatives such as DARPA's TRACTOR (Translating All C to Rust) program[35] to scale secure language migration for critical infrastructure.
- **Deep security tool integration:** AI code generation systems could incorporate static analysis, fuzzing, and other security tools into the generation and validation process (e.g., extensions of coding agents). This could provide real-time feedback and block unsafe suggestions by default.
- **Infrastructure-as-code security evaluations:** New evaluations specifically for AI-generated infrastructure configurations - Terraform, CloudFormation, Kubernetes manifests, and cloud provider settings - could assess whether AI systems produce secure-by-default configurations or introduce misconfigurations that create attack surface.

## Elements of success

Projects that successfully advance the security of AI-generated code will:

- **Define the target context:** Secure code generation projects should define the relevant context along key dimensions: whether the system targets novices or experienced security engineers; whether it operates as a fully autonomous agent or human-AI collaborative tool; whether it supports general-purpose development or critical infrastructure; and whether it aims to improve defaults in popular tools or build high-assurance specialized systems. The chosen context affects system requirements, evaluation methods, and where effort should concentrate. Contributors should define their target context early and reason through which security outcomes they aim to influence.
- **Prioritize ecosystem infrastructure:** In a fast-moving and increasingly competitive space, benchmarks, datasets, and evaluation frameworks provide the most durable value by creating incentives and infrastructure that benefit the entire field, including clearer signals about whether systems actually improve code security. Building highly specialized secure coding tools could have high impact on the security of AI-generated code, but they could also carry more risk. Well-resourced companies with frontier models may eventually outperform custom solutions unless there is a clear comparative advantage in domain expertise, proprietary data, or methodology.
- **Design for real-world adoption:** Projects should integrate into the environments where developers actually work, rather than assume theoretical or academic settings. They should plan for adoption from the outset: target integration into widely used IDEs and CI/CD pipelines, align with existing security tooling and code review processes, leverage open-source evaluation infrastructure where possible, and engage tool vendors and platform maintainers as partners.

# Develop assessment and auditing protocols for AI companies and AI systems

Importance #7

## What is this?

AI companies and AI systems face complex security risks that demand independent checks to confirm they are following strong security practices. To support these checks, stakeholders can develop practical methodologies, tools, and protocols to assess whether AI companies and systems meet security standards. Teams can perform red-teaming, purple-teaming, penetration testing, code and architecture reviews, automated compliance verification, and structured third-party audits. Government regulators, contracted auditors, industry certification bodies, and the organizations themselves can use these protocols to identify and mitigate AI-specific security risks across the development and deployment lifecycle.

## Why this matters

Assessing AI security requires practical methods to verify compliance, account for novel risks, and keep pace with rapidly evolving capabilities. The following points underscore the need for building effective assessment protocols.

- **Compliance with AI security standards requires practical, verifiable assessment methods.** Even as AI security standards and frameworks have proliferated (NIST AI RMF, ISO/IEC 42001, the RAND security levels framework, OWASP AI Security guidance, and others),[16,36,37] organizations and regulators still lack practical ways to verify whether organizations actually comply with them. Concrete assessment protocols can help auditors, regulators, and security teams confirm whether organizations are following established standards.
- **AI security risks present novel challenges that make assessments difficult.** While lessons from other domains - including aviation safety, nuclear regulation, and traditional cybersecurity[(2)38-40] - show that written standards alone cannot prevent failures, this gap is particularly acute in AI security, where many AI-specific risks - such as distillation and training data poisoning - do not fit existing audit procedures. The field also lacks consensus on what to test: reliable tools do not yet exist for many AI-specific attacks, and some assessment capabilities will require yet-to-be-achieved technical breakthroughs.

---

[2] For financial auditing, see the creation of the PCAOB under the Sarbanes-Oxley Act of 2002 in response to audit failures at Enron and WorldCom. For nuclear safety, see the NRC's post-Three Mile Island shift from compliance-based to risk-informed inspection, described in NRC, "Reactor Oversight Process," NUREG-1649, Rev. 6, 2016. For aviation, see U.S. Department of Transportation Office of Inspector General, "FAA's Oversight of Boeing 737 MAX," AV2020020, 2021.

The rapid pace of AI capability development means frameworks calibrated to today's systems risk becoming obsolete as new capabilities and attack vectors emerge.

- **The absence of assessment protocols creates regulatory, supply chain, and compliance risks.** The lack of agreed-upon AI security audit procedures means regulators lack ways to evaluate compliance while organizations lack ways to demonstrate it.(3)[41] Organizations deploying third-party AI systems cannot reliably assess upstream providers' security postures, creating blind spots throughout the supply chain. Without protocols testing for real-world security outcomes, compliance risks becoming a box-checking exercise where organizations optimize for metrics rather than materially improving defenses.(4)[42,43]
- **Developing robust assessment protocols requires adapting existing cybersecurity assessment methods to AI-specific risks**. While some aspects of such efforts are straightforward, such as adapting established penetration testing to cover AI infrastructure, others could prove challenging, including reliably assessing whether a model's internal behavior conforms to stated safety properties.

## Suggested projects

The methodologies for testing compliance with AI security standards remain underdeveloped. The following projects represent concrete opportunities to develop assessment and auditing protocols:

- **Government-led red-teaming of frontier AI companies.** Government-run adversarial red-teaming, drawing on concrete intelligence, can provide higher-quality assessments than commercial alternatives.[16] Projects in this area should develop structured engagement models for government-led or government-augmented red-teaming of frontier AI infrastructure, including protocols for scoping engagements, handling sensitive findings, and ensuring remediation. Continuous or near-continuous adversarial assessment may prove more valuable than periodic engagements.(5)[44]
- **Purple-teaming for AI security assessment.** Purple-teaming, in which offensive and defensive teams work together through coordinated attack simulation, detection testing, and joint remediation, yields richer findings than siloed red-teaming or blue-teaming alone, because defenders can immediately learn from and adapt to offensive techniques. Projects could develop structured, AI-specific purple-teaming methodologies, including protocols for scoping collaborative engagements, sharing findings between offensive and defensive participants in real time, and translating results into concrete security improvements.

---

[3] For example, U.S. federal agencies have begun developing procurement requirements tied to security assessments, and the EU AI Act reaches full enforcement for high-risk systems in 2026. See European Parliament and Council, Regulation (EU) 2024/1689 (AI Act).

[4] For examples of this pattern - where auditable metrics become targets and organizations optimize for the metrics rather than the underlying security - see Charles Goodhart, 1984. For application to compliance regimes, see Marilyn Strathern, 1997.

[5] As an example, assessments can be modeled on the Nuclear Regulatory Commission's (NRC) resident inspector program. See NRC, "Resident Inspector Program Background," 2023. The program places two or more NRC inspectors at each operating reactor site full-time, providing continuous oversight distinct from periodic visiting inspections.

- **Continuous security validation.** Projects could develop frameworks for continuous AI security validation rather than relying on point-in-time assessments. Continuous validation could complement periodic third-party audits with ongoing verification that security controls are functioning across the targeted organization.
- **Standardized audit protocols and assessment methodologies.** Stakeholders should develop detailed, repeatable auditing procedures for evaluating AI security postures that specify documentation requirements, technical testing procedures, interview guides, sampling strategies, and reporting templates. Protocols should test actual implementation, not just documentation. Sector-specific variants, such as for defense or healthcare, should include additional criteria tailored to each domain's risks and regulatory requirements.
- **Automated AI security assessment tooling.** New tools could partially automate assessment of AI security postures. Near-term options could verify access controls for model weights, detect infrastructure misconfigurations, detect issues in AI supply chains, and check compliance. Longer-term goals include static and dynamic analysis of agentic AI systems, automated verification of model behavior, and continuous vulnerability monitoring. While some tools can be adapted from existing cybersecurity tooling, others will require research breakthroughs that may take years.
- **Framework for ensuring auditing quality, independence, and institutional design.** Stakeholders should define and establish an institutional framework for AI security auditing that clarifies who is qualified to conduct assessments, what independence requirements apply, and how to prevent conflicts of interest.[6] Stakeholders should consider other analogous safeguards, such as separation of audit from advisory functions, assessor rotation, and external quality reviews. The specialized expertise required in AI systems, adversarial machine learning, and relevant threat models remains limited, making workforce development a prerequisite.
- **Validity and reliability research for assessment methodologies.** Researchers should conduct empirical studies to determine whether assessment protocols actually work in the AI security context: whether they identify real vulnerabilities, whether different assessors achieve consistent results, and whether organizations that score well actually experience fewer security incidents. This work should also examine whether poorly designed assessments create compliance fatigue or a false sense of security.[(7)38]

## Elements of success

AI security assessment spans diverse organizational contexts, threat models, and system types. Successful projects should:

---

[6] As an example, the Sarbanes-Oxley Act restricts auditors from providing consulting services to their audit clients to preserve audit independence. Sarbanes-Oxley Act of 2002, Pub. L. 107-204, §§ 201–203 (auditor independence requirements, partner rotation, and prohibition of non-audit services to audit clients).

[7] As an example, frameworks could draw on models like the NRC's triennial reviews of its own inspection program. NRC, "Assessment of the Reactor Oversight Process," conducted triennially to evaluate inspection program effectiveness. See NUREG-1649.

- **Distinguish between different assessors.** Protocols should account for both assessor and practitioner perspectives and address the different levels of access, expertise, and authority.
- **Ensure structural independence and resistance to capture.** AI security auditing should build in safeguards to ensure funders do not influence results, including assessor rotation, separation of audit from advisory functions, and government spot-check authority
- **Be scalable across organizational types.** Assessment frameworks should be scalable and modular to account for different organization types, with additional depth for the most sensitive or capable systems. Scalability should not come at the expense of rigor for organizations developing or deploying high-risk systems
- **Account for AI system diversity.** Foundation models in training, fine-tuned models in deployment, agentic systems with tool access, and AI infrastructure components each present different attack surfaces and warrant distinct assessment approaches. Protocols should reflect these differences while maintaining enough commonality to enable cross-system comparison.
- **Manage proprietary concerns without letting them become a veto.** While assessments should include safeguards to prevent the disclosure of proprietary information, proprietary issues should not limit scope of assessments, especially for high-risk systems. For example, appropriate safeguards could include cleared assessors, tiered disclosure protocols, and strict data handling.
- **Acknowledge current limitations.** Some aspects of AI security assessment are ready for standardization today; others remain unsolved research problems. Acknowledging limitations could advance transparency and trust and be preferable to falsely claiming the ability to assess things that cannot yet be measured.

## Develop a tamper-proof secure weight module for AI model weights

Importance #8

### What is this?

Some of the world's most capable attackers, including nation-states and well-resourced organizations, have the tools and resources to target and undermine the security of AI systems.[16] Current protections, such as confidential computing[45] and trusted execution environments (TEEs) that isolate memory on conventional processors, offer some security but were not built to withstand sustained attacks by highly capable adversaries.

One pathway to protect against sophisticated actors involves developing a **secure weight module (SWM)** - a proposed hardware device that stores AI model weights and performs all computation within a tamper-resistant, isolated boundary. Unlike TEEs, an SWM eliminates general-purpose software layers and exposes only a minimal, hardware-enforced API for inference and potentially other functions. An SWM could offer key security advantages, including physical isolation, cryptographic attestation, and strong resistance to information leakage.[8]

### Why this matters

As AI systems become increasingly integrated into core business operations, national security functions, and critical infrastructure, model weights face exposure to high-value theft, misuse, and unauthorized deployment. SWMs could provide hardware-enforced protection for model weights, addressing numerous challenges:

- **Dramatically reduced attack surfaces.** Securing model weights against top-tier adversaries is one of the hardest problems in AI security. Even the most advanced TEE deployments depend on a stack of dozens of components, any one of which, if compromised, can expose the weights. An SWM collapses this stack to a single purpose-built device with a minimal interface.
- **AI labs risk theft of their high-value intellectual property.** Training frontier models requires hundreds of millions of dollars in compute, data, and engineering. The resulting weights are the end-product of that entire investment. If stolen, an adversary can replicate the model's capabilities, bypassing the underlying research entirely. Theft can imperil model developers' businesses and countries' national competitiveness, as key adversaries who successfully steal model weights could erase potentially years of technological advantage and exacerbate race dynamics.

[8] This architecture could offer two potential advantages over TEE-based approaches. First, the attack surface would be substantially smaller. An SWM would have no general-purpose instruction set, no operating system kernel, no hypervisor, and no driver stack - only fixed-function hardware implementing a defined API. Second, because the device would be purpose-built and architecturally simpler relative to a commodity GPU, it may be more amenable to formal verification. Although likely impractical for commodity processors with billions of transistors and complex microarchitectures, a dedicated inference accelerator with a constrained instruction set is a potentially more tractable target. Together, these properties could make it possible to reason about the security of the device with greater rigor than is currently available for TEE-based approaches.

- **Unprotected model weights pose risks of systematic abuse.** Many safeguards in deployed AI systems, such as content and use-case limitations, often reside in the weights themselves or in tightly coupled parameter layers. Research has demonstrated that these safeguards can be removed with relatively few fine-tuning examples while retaining full capabilities.[46,47] If an adversary obtains the weights, they could strip safeguards and redistribute an unconstrained version of the model at scale. This differs from manipulating the model through input prompts, which could produce temporary unsafe behavior and does not allow the creation of a fully unconstrained copy of the model. Secure hardware custody could make this substantially harder, although it is not a complete solution on its own.
- **National and international security environments require strong model security.** SWMs could allow AI deployment in areas where TEE-level security is insufficient, such as in highly sensitive national security environments. They could also support international multilateral verification regimes, where hardware-enforced custody of weights provides a basis for verifiable commitments about model deployment and access.[48]

## Suggested projects

The following projects outline potential efforts that could advance the development and implementation of an SWM:

- **An extraction-resistant inference interface.** A key research need is defining and prototyping an interface for AI inference that bounds the rate of information leakage to acceptable levels (or prevents extraction entirely). Potential approaches include restricting output granularity (e.g., returning only top-k tokens rather than full probability distributions and withholding chain-of-thought traces), adding calibrated noise drawing on differential privacy or PAC privacy,[49] and monitoring query patterns to detect extraction attempts. This will likely require collaboration between hardware security researchers, ML theorists, and differential privacy experts, and may necessitate co-design of model architectures alongside the interface. It is analogous to the decades of cryptographic research that made HSM operations provably safe; no equivalent body of work yet exists for AI inference.
- **A highly-constrained inference-only hardware accelerator.** The SWM concept requires placing a general-purpose AI accelerator - effectively a full GPU - inside a tamper-resistant enclosure that only permits inference operations. This is substantially harder than existing HSMs, which contain relatively simple cryptographic FPGAs. This project would develop a dedicated inference accelerator for transformer-based models, enforcing all security properties at the hardware level. The accelerator would only perform specific operations needed for inference, using a limited set of instructions and no general-purpose commands that could be exploited to copy or extract the model weights. On-chip memory would be strictly divided between a write-once weight store and a separate area for inputs, intermediate values, and outputs, connected only through fixed pathways that prevent any weight data from leaking to the processing region. Additional architectural properties (temporal partitioning of bus access, pointer authentication on memory references, and hardware-enforced control-flow integrity) would further ensure that no timing-based side channels or invalid operations can leak weight data. A separate independent runtime integrity monitor would continuously verify that the chip is executing only valid operations and raises a fault if anything anomalous is detected.

- **Technical specifications for tamper resistance against OC5 (highly-resourced nation-state) adversaries.** Current standards like FIPS 140-3 provide a foundation for tamper resistance, but they were not designed with nation-state adversaries explicitly in scope. An SWM intended to secure frontier model weights against top-tier actors requires tamper-resistance specifications that go beyond FIPS 140-3, incorporating active intrusion detection, zeroization, hardened enclosures, and defenses against sophisticated physical, electromagnetic, and power-analysis attacks. Tamper resistance against state-level adversaries is an active area of research, with estimates of 18 months to 4 years of development from leading firms.[50]
- **A tamper-resistant GPU module (integrating the above).** The end goal is the integration of the above projects: a tamper-resistant hardware module containing a high-performance inference accelerator, exposing only an extraction-resistant API, with attestation and lifecycle management capabilities. This is the full SWM. It is likely several years away from production readiness.
- **Attestation, key management, and lifecycle protocols.** Remote attestation allows external parties to verify that an SWM is genuine and correctly configured before provisioning weights. Designing attestation schemes that remain robust across firmware updates and large-scale deployments is a core systems challenge. Key management (who generates, holds, and rotates the cryptographic keys used for attestation and weight encryption) has significant policy implications and must be addressed early. In TEE-based systems, key management defaults to the hardware vendor; an SWM ecosystem may require a different trust model.

## Elements of success

The success of SWMs will depend on key design and implementation factors. Successful projects will:

- **Ensure hardware-enforced guarantees.** The core custody properties (non-extractability of weights, restriction to a minimal API, tamper detection and response) must be enforced by hardware that cannot be bypassed by any software compromise. Relying on policy, configuration, or software controls alone is insufficient against advanced attackers. The device should be designed for formal verification of its critical security properties from the outset.
- **Achieve operational viability at scale.** SWMs must support large models without unacceptable performance degradation (e.g., targeting less than 10%). The primary performance bottleneck is likely the inference accelerator design itself rather than security overhead.

- **Maintain robust lifecycle management.** Secure patching, firmware updates, and key rotation must be possible without breaking attestation guarantees or creating windows of vulnerability. Changes to device firmware can affect attestation measurements, so lifecycle workflows need careful design to preserve trust continuity.
- **Ensure compatibility with existing infrastructure.** Since this is an SL-5-level security mechanism intended for the most sensitive deployments, full compatibility with all existing AI and cloud infrastructure may not be achievable or necessary. However, integration with major inference frameworks and deployment patterns would significantly increase adoption. Tooling for development, debugging, and structured diagnostics (such as signed error reports or constrained telemetry that does not leak weight information) will be important for practical use.
- **Mitigate risks of model extraction despite secure weight measures:** Even with SWMs, model outputs reveal information that attackers could use to reconstruct a functionally equivalent model through techniques such as model distillation, in which an adversary trains a separate model to replicate the target model's behavior using outputs obtained through query access. Addressing this limitation is a prerequisite for fully realizing the SWM vision and remains an open problem for many AI use cases. Successful projects should develop strategies and safeguards that limit this type of information leakage, thus addressing a key prerequisite for realizing the full potential of SWMs across AI use cases.

# Create specifications for top-tier AI security red-teaming

Cost-effectiveness #2

## What is this?

Comprehensive specifications for conducting elite-tier red-teaming exercises at AI labs could help prevent the theft of model weights[16] and algorithmic insights, as well as identify paths to sabotaging critical workloads. Specifications could include requirements for team composition, resources, attack permissions, issue resolution processes, and accountability mechanisms.

## Why this matters

While the limitations of commercial red-teaming affect many industries, AI organizations are distinctive in that the value of their core assets and the severity of the consequences if those assets are compromised may warrant defenses calibrated to the most capable adversaries. The following realities underscore why existing practices fall short and why more robust, standardized approaches are needed:

- **Existing red-teaming does not address the most advanced threats.** Red-teaming is a ubiquitous commercial security service, and many providers already offer red-teaming of AI systems. For most threat models - including persistent non-state actors (OC3) and some leading cyber-capable institutions (OC4) - commercial red-teaming is generally available, although often expensive. However, no commercial service can fully simulate highly-resourced nation-states (OC5). This gap provides little assurance against OC5 actors whose capabilities far exceed those of conventional red teams.
- **High-value AI assets face steep risks.** AI organizations, model weights, algorithmic insights, and training infrastructure represent - or could become - high-value targets for nation-states.[51] Without new, stronger red-teaming specifications, red-teaming could fail or provide a false sense of security, as organizations frequently under-resource teams, restrict attack permissions, or lack enforcement mechanisms for remediation.
- **The field lacks standardized specifications.** Organization-agnostic standards for elite red-teaming, calibrated to different threat models, do not yet exist, leaving AI labs without a reliable, externally verifiable way to evaluate defenses against OC5 threats.

## Suggested projects

Stakeholders could undertake a set of projects to develop adversary-calibrated red-teaming specifications that together address multiple aspects of assessment and mitigation. Ideally, these efforts would converge into a single, holistic framework that integrates adversary-calibrated requirements, OC5 threat assumptions, resolution and retesting protocols, and incentives and accountability measures. While each component could be developed independently, coordinating them in an integrated manner would substantially increase the practical value of the specifications and strengthen the rigor of elite red-teaming engagements. Projects include:

- **Development of adversary-calibrated specifications:** Stakeholders could develop new comprehensive frameworks that specify red-teaming requirements calibrated to different threat models, from OC3 (persistent nonstate actors) to OC5 (highly-resourced nation-states). These frameworks could define the appropriate team composition, resources, and attack permissions for each adversary model. For instance, specifications might define required expertise depth, team size, engagement duration, levels of red-team initial access to internal code, or the complexity of malware deployable during testing.
- **OC5 threat assumptions:** Effective OC5-level red-teaming requires clearly defined parameters that specify the red team's operational scope and capabilities. Scenarios should give red teams elevated capabilities to replicate OC5-level threat actors - including insider credentials, network access simulating zero-day exploitation, physical facility access, and white-box system knowledge - informed by realistic intelligence about adversary capabilities.
- **Development of resolution and retesting protocols:** Stakeholders could design frameworks for translating red-team findings into security improvements. Specifications could define severity-based remediation timelines (for example: while urgent vulnerabilities require two-week fixes, high-priority issues could require four-week fixes), as well as retesting requirements to verify that fixes are effective. They could also establish criteria for determining when broader architectural changes are needed beyond specific patches. While actual enforcement of remediation timelines will depend on organizational policies or regulatory requirements, embedding these recommendations within the red-teaming specification ensures they are addressed from the outset. Specifications should further clarify how to handle unexploited vulnerabilities - where red teams identify weaknesses but are unable to exploit them during engagement.
- **Incentives and accountability frameworks:** Stakeholders could develop criteria for selecting and pre-approving elite red-teaming organizations, design financial and reputational incentive systems that reward thorough vulnerability discovery rather than minimalist efforts, create mechanisms ensuring diverse attack strategies across exercises, and, for highest-assurance contexts, design external accountability measures as needed.

## Elements of success

The following elements of success outline key considerations for developing effective specifications for elite AI red-teaming. They highlight the design principles and operational factors that contribute to high-quality, credible exercises. Successful projects will:

- **Include concrete specifications about resources and permissions.** Stakeholders should avoid vague guidance - which could enable organizations to under resource exercises, impose unrealistic constraints, and claim compliance while remaining vulnerable - and instead establish clear, detailed guidance. Successful specifications explicitly state required investments - both in terms of person-years of elite expertise and operational budgets for their various needs. They should also clearly describe the aggressive tactics that can be critical for adequate red-teaming: running malicious code, conducting social engineering, deploying malware, utilizing real or simulated human intelligence, and testing physical security. At the same time, specifications should balance prescriptiveness and adoption; overly rigid requirements risk rejection from some organizations, while insufficient detail reduces effectiveness. Specifications should also address when teams could operate without security team knowledge to realistically test detection capabilities.

- **Ensure sufficient access and resources for OC5-level exercises.** Specifications should define how to safely share highly privileged information with red teams, compensate for the scarcity of OC5 expertise, and account for the substantial resources required to conduct thorough exercises.
- **Establish clear accountability for remediation.** This includes severity-based timelines, retesting requirements to verify fixes, and criteria for when findings indicate systemic issues beyond individual patches. Specifications should also address team selection criteria and incentive structures that reward thorough vulnerability discovery over minimalist compliance.
- **Address AI-specific attack vectors that traditional red teams may overlook.** This includes targeting model serving infrastructure, exploiting ML pipeline vulnerabilities, testing access controls around training data and checkpoints, and assessing whether compromised internal AI tools could accelerate an attacker's capabilities. Elite red teams for AI labs need expertise spanning both conventional security and AI systems - specifications should ensure team composition reflects this.

# Chapter 5: Governing agentic AI under adversarial pressure

As AI systems gain greater autonomy, governments, AI labs, and civil society face the challenge of maintaining meaningful human control over agentic AI - systems capable of acting, adapting, and interacting with the world in complex, often adversarial, contexts. Increasingly, these systems operate as active participants in security-critical environments rather than passive tools, concentrating autonomy, access, and adaptability in ways that existing governance approaches struggle to manage. because these systems now have real access to infrastructure and decision-making authority, these risks are no longer hypothetical, demanding practical and actionable risk mitigation strategies.

This chapter examines how to govern increasingly autonomous and capable AI as it evolves from passive tools into agentic actors that can plan, execute multi-step tasks, and modify their environments. We examine four priority areas that, together, could reinforce human control over agentic AI. Each represents a distinct approach - including governance, system design, technical safeguards, and adversarial threat modeling. These priority areas complement one another, as no single solution is likely to address the full spectrum of risks posed by agentic AI. These are:

1. **Permission frameworks** shape what agents can attempt
2. **Formal verification** defines which security properties must be guaranteed by construction
3. **Risk management frameworks** determine how autonomy is governed over time
4. **Threat modeling** clarifies what must be defended against.

These priority areas focus on system-level security rather than model-level behavior in isolation. Many of the most serious risks from agentic AI do not arise from individual model outputs, but from how models are embedded in systems with access to tools, data, and authority. We therefore emphasize architectural, governance, and control mechanisms that operate outside the model itself, shaping how agentic systems are deployed and constrained in practice.

## Develop permission frameworks for AI agent interactions

Importance #3

### What is this?

Permission frameworks represent potential best practices and software infrastructure to manage when and how AI can operate and interact autonomously in ways that carry material cost or risk. Projects could include designing auditable permission structures, building human oversight systems for sensitive actions, and creating shared protocols for granular AI access control.

### Why this matters

As AI agents take on a growing share of operational tasks, organizations will need to rethink how permissions and authorization are managed. Existing approaches designed for human users - including authentication, authorization, and accounting (AAA) frameworks - do not fully account for the scale, speed, and behavioral characteristics of AI agents.

- **If using existing frameworks, humans could spend significant time managing AI agent permissions.** Reliance on humans to manage approvals for AI agents could significantly increase the amount of time they spend on authorizing actions they did not initiate, making reviewer fatigue and rubber-stamping a systemic risk.
- **AI agents compound the security assurance challenge.** As agents produce and modify software, configurations, and infrastructure at increasing volume and speed, traditional review-based security assurance (such as peer code review to ensure services do not access unauthorized data) becomes a bottleneck. This dynamic creates pressure to adopt architectural approaches that enforce security properties structurally rather than relying on human verification of each change.
- **Traditional assumptions that underpin multi-party authorization (MPA) do not work for AI agents.** Simply delegating both an action and its review to separate AI agents provides weaker security guarantees than if using separate humans for these tasks. AI agents trained on similar data are likely to share systematic blind spots and may be susceptible to the same adversarial inputs, making correlated mistakes and shared vulnerabilities more likely.[(9)][52] Authorization frameworks that currently rely on MPA as a safeguard will need alternative mechanisms when agents replace human reviewers, since MPA's effectiveness depends on reviewers bringing differentiated context and making relatively uncorrelated errors.
- **Existing authorization frameworks do not address these dynamics.** Current identity and access management systems were designed for humans with stable roles and relatively predictable access patterns. AI agents that are short-lived and operate across organizational boundaries will likely need to be constrained differently, requiring new primitives that existing frameworks do not provide.

[9] For analysis of how correlated failures among AI systems may weaken monitoring-based safety strategies, see Greenblatt et al., 2024.

## Suggested projects

As development toward new AI-specific permissions frameworks remains relatively nascent, early contributions could prove particularly valuable. The following projects address different aspects of AI agent authorization, from adapting existing infrastructure to building new capabilities:

- **Existing infrastructure assessment.** Evaluating permission and delegation models, including fine-grained microservice authorization, OAuth extensions, delegation protocols, and identity frameworks, could help identify what can be adapted for AI agents and where AI-specific requirements (correlated failures, ephemeral identities, large action-context) demand novel approaches. This assessment could include a gap analysis documenting which existing primitives are sufficient, which need extension, and which AI-specific challenges require new solutions entirely.
- **Prototype high-stakes agent systems.** New prototype systems in contexts where the permissions pose substantial difficulties - such as agents accessing multiple enterprise systems, modifying sensitive data, or making financial transactions - could help identify practical requirements and failure modes.
- **Early deployment analysis and failure-mode cataloging.** Analysis could document how organizations currently handle AI agent permissions, informally catalog workarounds, and identify failure modes and the gap between intended policy and actual practice. The output could include a structured taxonomy of permission failures that informs standards development and architectural guidance, rather than a descriptive survey.
- **Open standards and shared infrastructure.** Open standards for AI permission management, including reference implementations, API specifications, and integration patterns, could address common patterns such as scoped credential delegation, action-context packaging for reviewer interfaces, and audit trail formats that support cross-organization interoperability.
- **Architectural trade-off guidance.** Experts could evaluate design choices such as cryptographic permission tokens versus centralized verification and determine how AI-specific characteristics could affect these trade-offs.[10]
- **Context presentation architectures.** Engineers could design systems that present relevant context to users when requesting permission for AI actions, addressing challenges such as large or complex information sets. Projects in this area could explore tiered context presentation (summary with drill-down), commitment mechanisms that allow reviewers to verify summaries against raw data, and interface designs that help reviewers identify the highest-risk aspects of a proposed action efficiently.

---

[10] For example, some cryptographic approaches (e.g., macaroons) avoid network round-trips but are harder to revoke. For AI agents, which typically operate on longer timescales than automated microservice-to-microservice calls, this tradeoff may be more favorable: revocation checks can be incorporated without unacceptable latency.

- **Speculative execution for authorization.** Engineers could design authorization systems that allow AI agents to proceed with tentative actions before receiving explicit approval, with mechanisms to roll back changes if permission is denied retroactively. This could prove particularly relevant for long-running agent workflows where blocking on every sensitive action would make the system impractically slow, but where full autonomy is infeasible. Key design challenges include defining which actions are reversible, how to bound the cost of rollback, and how to present action chains to reviewers.

## Elements of success

Effective AI agent authorization frameworks must balance security rigor with practical adoptability across diverse deployment contexts. Successful projects will:

- **Prioritize reviewer efficiency over agent convenience.** Authorization interfaces should make it easy for human reviewers to understand what an agent is requesting and to quickly assess whether an action could have costly consequences. As agent-driven workflows increase the volume of approval requests, the cost of reviewer attention becomes the binding constraint, and frameworks that do not account for this will either be bypassed or create unsustainable operational overhead.
- **Enforce ephemeral, scoped identities for AI agents.** AI agents should operate as ephemeral principals with task-specific permissions rather than inheriting the delegating user's full privilege set. This limits blast radius when an agent is compromised or behaves unexpectedly and makes audit trails more informative by tying actions to specific task contexts rather than broad user identities.
- **Reduce reliance on agent-to-agent review as a safeguard.** Given the risk of correlated failures among agents trained on similar data, authorization architectures should not depend on multi-agent review as a primary security mechanism. Human review, hardware-enforced constraints, or formal verification provide stronger independence guarantees.
- **Provide flexibility across trust contexts.** Frameworks must serve diverse deployments: high-trust internal agents versus public-facing agents requiring strict oversight, real-time interactions versus longer tasks where verification is acceptable, and stable permission environments versus rapidly evolving policies. A single core infrastructure should scale across these contexts without requiring fundamentally different architectures.
- **Design for change and auditability.** Permission structures will evolve as agents are deployed more widely and as regulations develop. Systems must support versioned policies while maintaining the ability to audit actions taken under previous permission regimes, ensuring that historical accountability is not lost as policies are updated.
- **Compose with existing security infrastructure.** The most adoptable frameworks will integrate with existing identity and access management systems while adding primitives for AI-specific concerns such as action explanation, graduated escalation for ambiguous situations, and context summarization for reviewers. Frameworks that work alongside both conventional security tools and emerging AI safety measures will have the broadest applicability.

## Establish formal verification of security properties

Importance #9

### What is this?

Formal verification of security properties provides guarantees that AI systems and the infrastructure that supports training, research, and deployment behave exactly as intended. For instance, verification can ensure that an AI agent can only operate within approved systems, that sensitive processes remain fully isolated and secure, that access controls strictly govern which resources a process can read or write, and that limits on data exfiltration are reliably enforced.

### Why this matters

As AI agents become more capable, their deployment in sensitive contexts will introduce new security challenges. Ensuring that these systems behave safely and reliably under all conditions will require new approaches to securing the infrastructure that supports them. Key considerations for why formal verification matters in these contexts include:

- **AI agents may act in ways that humans cannot fully monitor.** As agents become numerous, fast, and highly capable, it will likely be infeasible for humans to observe a substantial fraction of their actions, creating opportunities for malicious behavior to go unnoticed.
- **AI agents might attempt to exploit vulnerabilities in their deployment environments.** The underlying models could develop misaligned objectives and autonomously work to undermine their developers and users.[52] Such exploitation could also occur if human adversaries compromise them, for example, through their training data ("data poisoning").
- **Formal verification provides strong, mathematically proven security guarantees.** Security mechanisms backed by formal proofs could help defend against these threats. Formal verification provides guarantees that certain security properties hold under all possible conditions, eliminating entire classes of vulnerabilities rather than patching individual instances.
- **Formal verification cannot eliminate all risks, but it can secure substantial portions of vulnerable components.** Even securing only those aspects that can be formally defined substantially reduces the ways AI agents could exploit the system.
- **The emergence and improvement of AI-generated code could make formal verification practical at scale.** Historically, few software engineers could use formal verification tools, and producing verified software was extremely labor-intensive. Advances in AI coding agents may allow formal verification to be applied more broadly and at lower cost, enabling many infrastructure components to be provably secured that were previously infeasible.

## Suggested projects

Key projects in this space include:

- **Data exfiltration bounds**: Verified network infrastructure could help guarantee that a network node, process, or AI agent cannot upload more than a specified amount of data to external networks.
- **Verified interface boundaries**: When multiple AI agents collaborate on shared infrastructure, formal verification could help to create programming environments that enforce strict interface boundaries that cannot be violated. Creating verified interface boundaries could help ensure that understanding one component's behavior requires only understanding its specified interfaces, not implementation details of other components. Most current environments do not enforce this property. Achieving it would require formal verification of compilers or programming language runtime systems.
- **Hypervisor and network verification**: Building formally verified systems could provide certainty about information flows between computers in a data center, including provable guarantees about network configurations and routing. For example, in the context of AI deployed internally within AI companies, this work can leverage the fact that internal AI deployments require less confidentiality from the datacenter operator than typical datacenter scenarios, enabling verification approaches (such as traffic inspection) that would not be feasible in traditional multi-tenant environments.
- **Bounded information flow for resource sharing**: Developing verified mechanisms could allow safe resource sharing between AI agents while provably bounding the rate of information transfer. This will be particularly important if organizations shift from inefficient isolation (separate Docker containers, separate machines) toward shared infrastructure where multiple AI agents work on the same systems.
- **Behavioral equivalence verification**: Creating tools that formally verify an AI's reimplementation of software could help to ensure that the AI maintains the original's functional behavior and security properties, allowing performance optimization without introducing security regressions.
- **Declarative system specifications**: Stakeholders could develop frameworks that allow system configurations and security properties to be specified declaratively rather than imperatively. These frameworks could let operators define desired states, such as "agent A can communicate with agent B but not agent C," rather than writing step-by-step code to set up network routing, firewall configurations, or access control policies. Declarative specifications are typically easier to formally verify, and audit. Formal verification could then ensure that an efficient implementation actually ensured the properties written in the declarative specification.
- **Verified monitoring and audit systems**: Stakeholders could build formally verified logging and monitoring infrastructure that provides guarantees about the completeness and integrity of audit trails.

## Elements of success

Successful formal verification projects for AI control must navigate several competing pressures. The most impactful work will identify specific, high-value security properties where formal verification is both critical for risk reduction and becoming newly tractable through AI-assisted development. Successful projects will:

- **Include practical deployment requirements**: Verification systems must not be so restrictive that they prevent productive AI use. Organizations will deploy AI agents extensively, so security measures need to enable them rather than block legitimate work.
- **Leverage AI deployment context**: Internal AI deployment differs from typical security scenarios: privacy concerns may be reduced, labor costs for verification are lower, and organizations have more control over the entire stack. Successful projects should exploit these differences rather than merely adapting solutions from other domains.

# Develop a risk management framework for agentic AI

Importance #10, Cost-effectiveness #4

## What is this?

A Risk Management Framework (RMF) for agentic AI is a reference framework that provides a structured set of lifecycle activities, roles, decision points, and controls. Organizations can adopt or adapt RMFs to manage risks from AI systems that plan and execute multi-step actions using tools, memory, and ongoing interaction with environments. Rather than a new government compliance standard, RMFs could serve as practical blueprints that map agentic-specific risks to established risk-management functions (e.g., scope definition, risk assessment, controls, monitoring, accountability, governance). An agentic-AI RMF would set clear autonomy boundaries, enable real-time oversight, and ensure timely detection and response to behavioral deviations before they lead to unacceptable outcomes. This approach builds on emerging guidance from NIST, OpenAI, Anthropic, and Google DeepMind, which emphasizes continuous monitoring and capability-aligned governance.[28,53-55]

## Why this matters

As agentic AI systems gain autonomy and operate in complex, dynamic environments, organizations face new challenges in overseeing their behavior. Key factors highlight the value of a risk management framework:

- **The ability of agentic AI to evolve after deployment creates new risks - and warrants a new governance approach.** Agentic AI systems warrant adaptations to existing risk-management practices, because their ability to plan and act overtime introduces risk dynamics that can evolve after deployment. Traditional frameworks such as NIST's AI Risk Management Framework (2024) and ISO/IEC guidance provide strong foundations, but they offer less specificity on how to govern ongoing, stateful decision-making in changing environments (e.g., tool use, memory, autonomy escalation, and multi-step execution). As a result, organizations lack tools to manage the ongoing behaviors that determine whether an agent remains safe and aligned.
- **An RMF can help provide continuous oversight and identify early warnings.** An RMF for agentic systems would require continuous oversight of planning, action, and adaptation so that governance keeps pace with their evolving behavior. Research by Nichols, Nevo, and Greaves (2025) highlights how vulnerabilities can arise in memory management, communication pathways, and multi-agent coordination. These system-level factors can amplify small errors or create new failure modes even when models behave as intended in non-agentic settings. Without a framework designed for these dynamics, organizations may miss early signs of unsafe agentic behavior. An agentic AI RMF addresses this gap by establishing mechanisms for real-time monitoring, provenance tracking, and timely intervention when agents diverge from intended goals. These mechanisms ground accountability in measurable processes and enable organizations to detect and correct unsafe actions before they escalate into systemic failures.

## Suggested projects

To operationalize an RMF for agentic AI, several complementary projects could be pursued to develop, test and embed the framework in practice. These include:

- **Prototype RMF for agentic systems:[11]** This core project would produce a **prototype risk management framework artifact** that provides a structured set of lifecycle processes, roles, decision points, and controls. The prototype would be tailored to the distinctive behavior of agentic AI, rather than building isolated components in sequence. Standards are most effective when designed as an integrated whole to ensure coherence across autonomy boundaries, tool integrations, memory persistence, and escalation protocols. Drawing on ISO 42001 and OECD's reporting guidance, the RMF could incorporate continuous-improvement cycles and map policy decisions to measurable indicators and controls.[56,57] Current Responsible Scaling Policies (RSPs) from leading AI labs, such as OpenAI, Anthropic, and Google DeepMind, offer thresholds that could inform governance triggers and escalation criteria.[28,53,55]
- **Scenario-based evaluation and oversight systems:** Complementary projects could create the infrastructure needed to evaluate agentic AI systems in simulated settings before deployment and to oversee their behavior during live operation, ensuring that RMF requirements are met across the full system lifecycle. Scenario-based evaluation environments would test agents under conditions such as conflicting goals, delayed feedback, or adversarial inputs to assess their long-term planning and adaptability. Techniques from OpenAI's capability evaluations and Anthropic's AI Safety Levels could help benchmark risk-prone behaviors.[53,55] In parallel, runtime monitoring and intervention systems could build on OWASP's emerging guidance and the CSA's MAESTRO framework to detect irregular behavior, log agent decisions, and enable containment actions such as privilege revocation or execution halts.[37,58] Together, these evaluation and oversight systems would provide the evidence and controls needed to support the RMF's monitoring and validation functions.
- **Capability-threshold governance and accountability mapping:** A complementary effort could define capability-threshold triggers and accountability artifacts specified by the RMF, including when new agent abilities (e.g., modifying production systems, accessing external APIs, or creating sub-agents) require formal review, audit, or safeguard upgrades. This approach reflects practices described in frameworks developed by OpenAI, Anthropic, and Google DeepMind[28,53,55] and would further the conversation between the RMF and existing RSPs. Accountability and provenance tracking would define schemas for capturing how agent decisions are authorized and supervised, aligned with OECD's risk-reporting template and NIST's emphasis on traceability.[54,57] These records would enable post-incident analysis and continuous policy improvement, and they provide concrete artifacts for RMF audits and governance reviews.

---

[11] To conserve space, Table C.1 (see Appendix C) outlines notional RMF components that align agent-specific oversight processes with general risk-management functions.

## Elements of success

For an agentic-AI RMF to be implementable, auditable, and durable in real-world operations, it must translate high-level risk principles into enforceable technical and organizational mechanisms. Projects that do the following could help make the RMF effective in practice:

- **Ensure compatibility with enterprise risk and governance standards:** An effective agentic AI RMF should align with established risk management standards while extending them to address autonomous behavior. Frameworks such as NIST's AI RMF and ISO 42001 can be adapted to incorporate trajectory oversight, multi-actor decision-making, and dynamic authority models such that organizations can integrate agent-specific safeguards into existing enterprise-wide governance structures rather than creating isolated or duplicative controls.[54,56]
- **Establish clear, enforceable autonomy and authority boundaries:** A successful agentic-AI RMF begins with explicit constraints on autonomy. Organizations must define which systems and data an agent may access, which classes of decisions require human approval, and how authority may escalate or contract over time based on performance, context, or risk signals. These boundaries should not be merely policy-level declarations; they must be enforced through technical controls such as permission gating, tool-use restrictions, and intervention thresholds that reliably limit agent behavior in practice.[28,55]
- **Conduct trajectory-aware, process-oriented risk assessment:** Traditional output-focused evaluation is insufficient for agentic systems. Risk assessment must instead examine the full decision-making process: how agents generate and revise plans, select tools, modify their environment, and update internal state over extended action sequences. This procedural view of risk aligns with emerging frameworks such as CSA's MAESTRO and the CLTC risk profile, which emphasize understanding how outcomes are produced rather than evaluating outcomes in isolation.[58,59]
- **Perform continuous and adaptive monitoring and intervention:** Because agentic risks can emerge during deployment, RMF controls must be paired with monitoring that provides ongoing visibility into agent behavior (actions, tool calls, state changes) and predefined responses when thresholds are crossed (containment, de-escalation, privilege reduction, rollback). Monitoring intensity should scale with capability and operational context, consistent with tiered governance approaches such as Anthropic's AI Safety Levels framework and OpenAI's Preparedness Framework.[53,55]
- **Ensure auditable end-to-end traceability and accountability:** Agentic systems must support comprehensive traceability across their lifecycle. Each decision or action should be attributable to a specific configuration, authorization state, training version, and supervisory context. This level of traceability enables post-incident analysis, compliance verification, and continuous improvement. Recent updates to NIST and ISO guidance underscore the importance of lifecycle visibility as a foundation for effective risk management and mitigation.[54,56]
- **Include scalable governance that evolves with agent capability:** Finally, RMFs must be designed to evolve alongside agent capabilities. As systems gain longer planning horizons, broader tool access, or multi-agent coordination abilities, risk controls must scale accordingly without becoming operational bottlenecks. Success requires balancing safety, accountability, and organizational agility, ensuring that governance mechanisms remain effective as agentic AI systems transition from experimental deployments to core operational roles.

# Use threat modeling to understand risks of AI attacking its host infrastructure

Cost-effectiveness #5

## What is this?

Threat modeling AI as an attacker involves analyzing how AI systems - particularly autonomous agents with broad system access - could be used to attack the infrastructure that hosts them, at the direction of an attacker. The attacker may be a human adversary, such as an insider threat, cybercrime organization, or nation-state APT, or it could be another AI system. Attacks can occur through explicit adversarial instructions, emergent behaviors triggered by manipulated tasks, or exploitation of misaligned objectives.

**Note:** the vast majority of work on this would also have benefits if an AI model started behaving as an attacker even without a separate threat actor directing it to do so. Other work relevant to that scenario can be found in “Develop Permission Frameworks for AI Agent Interactions.”

## Why this matters

AI agents are rapidly transitioning from simple assistants to autonomous actors that execute complex, multi-step tasks with minimal human oversight. Understanding how AI systems could be used as attackers is critical to designing safeguards and protecting the infrastructure that hosts them.

- **AI agents can act as sophisticated internal adversaries.** When an AI system can write and execute code, browse the web, access internal tools, and communicate with other systems, it could compromise the very infrastructure that hosts it. Unlike traditional vulnerabilities, an attacker who can influence an AI agent gains access to a sophisticated actor already positioned inside the target environment. In some cases, this actor has vast pre-existing permissions and has the ability to act within private networks without arousing suspicion; or at the very least, might be able to pick courses of action without needing to communicate with the attacker via command-and-control (C2) channels.
- **Indirectly influenced behavior can lead to attacks.** Research has demonstrated that AI agents performing benign tasks can engage in offensive cyber behavior when facing obstacles - exploiting vulnerabilities, disabling security defenses, and developing exfiltration techniques to bypass data loss prevention.[60] The research indicates that this can even occur without explicit malicious directions; instructions to be creative, handle errors, and ensure task completion could enable attacks.
- **AI agents combine insider and outsider threat characteristics, further compounding the risks.** AI systems deployed internally combine "the hard parts of both" insider and outsider threats.[61] Like insiders, AI agents receive broad access with granular permissions, while, like outsiders, they could act at significant scale and potentially overpower labor-intensive insider security techniques. Traditional security frameworks are ill-suited for such a combination, as they often assume either numerous but restricted users or trusted but few privileged actors.

- **The threat grows with multi-agent deployments.**[62] As organizations deploy teams of collaborating agents, the attack surface expands - an attacker who compromises one agent can potentially propagate their influence to other agents. Understanding AI-as-attacker threat models now, while deployments are still scaling, will enable more effective security architecture decisions and incident response capabilities.

## Suggested projects

Preparing for AI agents as potential attackers will require systematic analysis, experimentation, and evaluation. Potential projects include:

- **Comprehensive taxonomy of AI agent attack vectors.** Stakeholders could compile a systematic catalog of the ways AI agents could attack infrastructure, distinguishing capabilities AI agents possess that humans do not (and vice versa).[63] For example, AI agents can operate at machine speed, coordinate with copies of themselves, and execute thousands of equally autonomous and intelligent parallel probes without fatigue. However, they lack physical presence: they cannot plug in a USB device, tailgate through a secure door, or read a password written in a notebook without human assistance. The taxonomy should map attack vectors by access method (e.g., exploiting developer credentials, persuading humans to grant access, direct exploitation) and attack types/goals (e.g., data exfiltration, privilege escalation, denial of service, etc.).
- **Threat models differentiated by threat actor and organizational context.** This project would analyze how different threat actors might direct or exploit AI agents to attack the infrastructure hosting them. For instance, while a human insider might use an AI agent to amplify capabilities or provide deniability, a cybercrime organization might compromise agents through prompt injections to establish persistent access. The attacker might itself be an autonomous AI system interacting with victim organizations' agents. Because each threat actor implies different threat models and mitigations, projects should develop modular threat-modeling frameworks customizable to different contexts - AI labs, government agencies, or enterprises.
- **Red-team exercises and controlled experimentation.** Stakeholders could develop controlled environments where security teams can experiment with adversarial AI scenarios. These environments would allow testing of situations such as agents receiving instructions designed to trigger offensive behavior, multi-agent scenarios in which adversarial agents attempt to compromise legitimate ones, and collaborations between humans and AI agents. These exercises should produce findings about which controls are effective and which architectures are more susceptible to manipulation.
- **Evaluation frameworks and benchmarks.** This project would develop standardized assessments to evaluate the susceptibility of different AI systems to adversarial manipulation. Benchmarks could measure resistance to prompt injections, propensity for offensive behavior under task pressure, effectiveness of guardrails, and security properties of multi-agent protocols. These assessments would allow organizations to compare configurations and give developers concrete targets for improving security.

## Elements of success

Effective threat modeling for AI agents requires a coordinated approach that draws on diverse expertise, anticipates novel behaviors, and adapts to evolving capabilities. Successful projects will:

- **Foster interdisciplinary collaboration between ML and cybersecurity experts:** Stakeholders should ensure projects involve both ML and cybersecurity practitioners to leverage expertise from both domains. ML researchers understand how language models behave - emergent capabilities, failure modes, architecture differences - and think in terms of evaluations, benchmarks, and scaling laws. Cybersecurity practitioners bring expertise in concrete system analysis, attack surface identification, and proofs-of-concept that demonstrate vulnerabilities. Projects lacking either perspective risk producing incomplete threat models: too abstract to be actionable, or too narrow to anticipate how threats evolve.
- **Include threat models that address both universal and context-specific concerns:** Projects should include threat models that combine general and context-specific insights, as AI labs, enterprises, and government agencies have different needs. Projects should aim for layered approaches: foundational threat models capturing universal concerns, combined with context-specific extensions addressing particular deployment scenarios and threat environments.
- **Address multi-agent and human-agent dynamics:** Multi-agent systems and agent-human teams create novel attack surfaces: adversarial agents infiltrating teams, agents manipulating other agents or humans, and cascading compromises where one breach enables others. Threat models should consider how trust forms, how malicious behavior propagates, and how control might be subverted.
- **Evolve with AI capabilities:** Work should grapple with uncertainty about future AI capabilities, since threat models developed today might become obsolete as systems advance. Successful projects should build in mechanisms for ongoing revision: living documents, extensible benchmark suites, and frameworks designed to accommodate uncertainty.

# Chapter 6: Conclusion

AI security remains an emerging field. Much of the institutional infrastructure, shared practices, and coordination mechanisms that underpin other security domains have yet to develop in the AI context. As a result, governments, researchers, companies, and individuals might struggle to determine where their efforts would matter most or how to begin contributing.

The expanding landscape of AI security risks demands urgent attention from stakeholders across the AI ecosystem - those already engaged and those newly joining the effort. This paper outlines a practical agenda to improve the state of AI security and address these emerging risks. The agenda includes 18 high-value priority areas - selected for their importance and cost-effectiveness - across four themes: (1) strategic foundations and policy frameworks, (2) public-private coordination, (3) technical security engineering, and (4) governance of agentic AI. Within each priority area, we identify a range of projects that readers could take on to meaningfully advance progress. We also define elements of success to help ensure these efforts succeed in practice.

By identifying concrete, expert-vetted priorities, this agenda highlights areas where work is both urgently needed and ready for meaningful progress, inviting new entrants and new ideas to shape the AI security landscape.

**This agenda represents only an initial, "living" draft.** As AI continues to advance, so too will the broader environment in which it operates: regulations and policy responses, knowledge of how AI works, and the nature of threats and vulnerabilities will all shift. Frontier AI security will require ongoing dialogue and iterative adaptation across sectors to build a security foundation that keeps pace with technological change.

By pursuing the priority areas outlined in this agenda - and by expanding and refining them as the field evolves - the AI ecosystem can take concrete steps toward securing increasingly powerful AI systems while preserving their potential to deliver long-term societal benefits. The work ahead is substantial, but it is also an opportunity: to build the foundations of a new field dedicated to ensuring that AI remains secure, resilient, and worthy of public trust.

# Appendix A: Methods

This appendix describes the methodological approach used to identify, prioritize, and analyze potential efforts in AI security. The study relied on four primary components: (1) initial expert interviews to surface a broad range of ideas and challenges in the field, (2) a workshop on AI security priorities to refine and collectively assess those ideas, (3) post-workshop effort-scoring interviews to estimate the practical difficulty of implementing each project, and (4) desk research to assess the level of existing activity in each area. These data streams were integrated to generate a set of priority topics and cost-effectiveness estimates, which informed the selection and authorship of eighteen priority areas.

## Initial interviews

Irregular and RAND researchers conducted five initial scoping interviews (of 30 invited) with experts recruited through professional networks and outreach to organizations active in AI governance, cybersecurity, and AI system development. Participants were selected for senior-level experience (at least five years) in one or more of the core domains relevant to this study: policy development for AI and information security, AI system design and deployment, cybersecurity, and research examining how AI systems influence information-security environments. Participants' sectors and primary domains of expertise are summarized in Table A-1.

**Table A-1. Initial Interviews: Participant Characteristics**

| Participant | Sector | Primary Domains / Expertise |
|---|---|---|
| Participant 1 | U.S. Government | Cybersecurity; AI |
| Participant 2 | UK Government | Cybersecurity |
| Participant 3 | Industry | Cybersecurity |
| Participant 4 | Nonprofit / Independent | AI policy |
| Participant 5 | Nonprofit | National Security |

Note: Interviews were conducted for early-stage project scoping rather than representative sampling. Participants were selected to ensure coverage of relevant sectors and expertise areas, and several perspectives informed discussions despite the small number of interviews.

Interviews were conducted via video conference and lasted approximately 60 minutes. Prior to implementation, researchers piloted a semi-structured interview guide internally to ensure clarity and relevance. Verbal consent was obtained at the start of each interview. Sessions were not recorded and no transcripts were produced; instead, researchers relied on structured notes taken during and immediately after the conversations. Because this phase served a scoping purpose, no formal qualitative coding was conducted. Two researchers reviewed notes to synthesize emerging themes and identify recurring concepts across interviews, which ultimately informed the preliminary list of 60 potential AI-security projects.

## AI Security Priorities Workshop

The AI Security Priorities Workshop convened 14 experts from industry, government, and civil society to assess and prioritize a broad set of AI security projects (see Table A-2 for participant characteristics). Invitations were issued through individual outreach and professional networks, with the goal of achieving a balanced distribution across sectors and areas of expertise. Participants were selected based on substantial experience in AI policy, national security, cybersecurity, AI development, or combinations of these domains. Participation was voluntary and uncompensated.

The workshop followed a structured, three-stage process: (1) brainstorming, (2) consolidation and deduplication, and (3) importance scoring and discussion. Prior to the event, participants received a pre-read packet containing the agenda, background materials, the initial project list, and instructions for the importance-scoring activity. No pre-work was required. Breakout groups were assigned to ensure diversity of sector and expertise representation, and each room included a trained facilitator and a designated notetaker following a consistent facilitation guide. The event was conducted under the Chatham House Rule, with verbal consent obtained at the outset.

In the first stage, participants engaged in structured brainstorming to propose additional project ideas beyond the initial list. In the second stage, researchers consolidated and deduplicated these submissions to produce a unified set of candidate projects. In the third stage, participants independently rated each project for importance, then discussed their reasoning in small groups. Missing scores were permitted and excluded from aggregation. Group-level importance scores were calculated as arithmetic means. Table A-2 summarizes the final sector and expertise composition of attendees.

**Table A-2. AI Security Priorities Workshop: Participant Characteristics**

| Category | # of Participants |
|---|---|
| **Sector**[12] | |
| Industry | 7 |
| Civil Society | 3 |
| Government | 4 |
| **Expertise Areas**[13] | |
| Cybersecurity & AI | 8 |
| AI Policy & National Security | 3 |
| Dual Expertise (both areas) | 3 |
| **Total Participants** | 14 |

## Post-workshop effort-scoring interviews

Following the workshop, seven additional experts participated in individual effort-scoring interviews conducted via video conference. Participants included individuals who had been invited to the workshop but could not attend, as well as additional experts selected for their ability to estimate the scale of effort required for each project. Approximately 20 invitations were sent. All participants had experience in AI and cybersecurity, and four also had experience in AI policy and national security.

The effort-scoring process followed a standardized protocol consisting of three steps: (1) orientation and rubric review, (2) live scoring and clarification, and (3) score finalization. Each session lasted approximately one hour. Participants completed the scoring live using the provided rubric, with facilitators available to clarify definitions or criteria but not to offer substantive judgments. Scores remained private and accessible only to the research team. Participants could revise their ratings before finalizing their assessment. Missing scores were permitted and excluded from aggregation. Researchers recorded notes during each session.

Effort scores were aggregated as arithmetic means, with no adjustments or outlier corrections applied.

## Desk research on existing efforts

Desk research was conducted to estimate the number of existing efforts addressing each priority area. The process followed three main steps: (1) source identification and search design, (2) effort screening and inclusion, and (3) data consolidation.

In the first step, two researchers reviewed publicly available sources, including academic literature databases, government and intergovernmental reports, think-tank publications, company blogs, GitHub repositories, arXiv submissions, conference proceedings, and general web search results. Searches used both priority-area-specific and general AI security–related terms and focused on work from approximately the past five years.

[12] Sector distribution reflects participant records from the workshop sign-in and registration process.

[13] Expertise categories are based on self-reported backgrounds and pre-workshop designation; participants could be included in more than one category.

In the second step, researchers applied two criteria to determine whether an item qualified as an existing effort:

1. The effort must be specifically directed at the priority area, and
2. It must demonstrate appropriate scale for the project.

For this study, "appropriate scale" referred to efforts that could be expected to make substantial progress on a given project. Student projects, hackathon demonstrations, conceptual pieces without implementation, and casual organizational references were excluded.

In the third step, each researcher independently searched for efforts across all projects, then collaboratively adjudicated duplicates and disagreements about exclusions. Final findings were recorded in separate spreadsheets.

## Integration and cost-effectiveness calculation

Importance, effort, and existing initiatives data were combined to produce a cost-effectiveness estimate for each project. All projects had complete data, and none were excluded. Importance, effort, and existing initiatives were treated independently during data collection and merged only during the cost-effectiveness calculation.

Importance scores were normalized to better reflect the sense of importance provided in the scoring guide. The five-point importance scale (1–5) was remapped to –10, 0, 1, 10, and 100, based directly on the wording provided to participants in the scoring guide. The remapping reflects the qualitative descriptions associated with each score, as outlined in the workshop documentation.

The cost-effectiveness estimate was calculated as:

$$CE = \frac{Average\ Importance}{Average\ Effort\ x\ (Existing\ Efforts + 1)}$$

The addition of "+1" implements a counterfactual adjustment: given the number of other pre-existing efforts aimed at the projects goal, and with no information on the relative capacity or contributions of the different efforts, the expected contribution of any single effort is estimated as 1/(existing efforts +1), dividing value equally among all the efforts, and including +1 for the additional project that would be hypothetically set up.

## Measuring expert disagreement on importance scores

The nonlinear remapping of importance scores introduces a complication for measuring expert disagreement, which this section describes along with the approach used to resolve it. As described above, importance scores were remapped from the original five-point scale (1–5) to a nonlinear scale (–10, 0, 1, 10, 100) designed to better reflect the qualitative descriptions provided to participants. This remapping complicates the interpretation of standard deviation as a measure of disagreement among experts. On the nonlinear scale, the interval between scores of 4 and 5 (mapped to 10 and 100, respectively) spans 90 points - approximately 82 percent of the total scale range - while the intervals between all other adjacent scores are far smaller. As a result, standard deviations computed on the remapped scale are dominated by variation between scores of 5 and non-5, producing values that appear disproportionately large relative to the mean. For example, in a hypothetical panel of ten, a project rated 4 by seven experts and 5 by three experts, reflecting modest disagreement on the original scale (standard deviation of 0.48), yields a standard deviation of 43.5 on the remapped scale, comparable in magnitude to the mean itself. This effect is not a computational error but an inherent property of any scale with this degree of nonlinearity.

Computing standard deviations on the original 1–5 scale avoids this distortion but creates a unit mismatch with the remapped means, making the values difficult for readers to interpret.

To address both problems, the main text presents expert disagreement using a qualitative color scale rather than a numerical standard deviation. Standard deviations were first computed on the original 1-5 scale, where values ranged from 0.50 to 1.34 across the 68 projects, with an average standard deviation of 0.89. These values were then mapped linearly to a color gradient ranging from white (standard deviation of 0, indicating full agreement) to red (standard deviation of 2, the theoretical maximum on a five-point scale, achieved when respondents are evenly split between the endpoints). Intermediate shades of pink indicate moderate disagreement, with deeper color corresponding to greater dispersion. The resulting disagreement indicator appears as an additional column in the expert ranking tables in Appendix B. This approach conveys the degree of expert consensus in an intuitive visual form while avoiding the interpretive difficulties associated with either numerical representation.

For transparency, **Table A-3** reports both the standard deviation on the original 1-5 scale and the standard deviation on the remapped –10 to 100 scale for each project.

**Table A-3. Priority Area Full Standard Deviation Scores**

| Priority Areas | Cost Effectiveness | Importance | 1–5 Standard Deviation | –10 to 100 Standard Deviation |
|---|---|---|---|---|
| AI Security resources (usable documents and training materials): set up a hub (e.g., public git) of AI Security resources, including materials on defending from nation-states | 2.29 | 8.75 | 1.06 | 27.82 |
| AI security posture assessment: develop assessment/auditing protocols for AI companies and AI systems | 1.74 | 36.83 | 1.1 | 44.83 |
| Elite security red-teaming specifications: create specifications for a top-tier AI security red-teaming effort | 1.27 | 18.00 | 1.22 | 35.40 |
| Incident response: develop AI-specific incident response recommendations | 1.00 | 11.50 | 0.69 | 26.96 |
| Risk management framework for agentic AI (as opposed to RMFs for foundation models) | 0.84 | 33.14 | 0.82 | 42.45 |
| Threat Modeling AI as an attacker: at higher capabilities, the model may attack the infrastructure that is hosting it autonomously (at the direction of an attacker). | 0.65 | 29.13 | 0.83 | 41.10 |
| Incident reporting: develop recommendations for protocols for sharing AI security incidents | 0.59 | 32.50 | 0.86 | 42.86 |
| AI security standards: formulate technical standards for securing the critical components of AI models | 0.54 | 32.50 | 0.82 | 42.86 |
| Implications analysis: develop a paradigm of offensive cyber operations to assess which capabilities lead to which societal outcomes | 0.53 | 13.18 | 0.84 | 27.79 |
| Taxonomy and language clarification: Improving our taxonomy of "AI security" and how it involves and relates to trustworthiness/control/alignment | 0.51 | 15.20 | 0.97 | 28.62 |
| Governmental support for the security of AI labs: Mechanisms for background checks for sensitive personnel | 0.48 | 24.20 | 1.13 | 38.14 |
| National AI deterrence strategy: USG clarifying strategic posture to deter other nations' theft or sabotage (e.g. committing to sanctions in response to model weight theft) | 0.18 | 46.89 | 1.12 | 47.63 |
| Governmental support for the security of AI labs: Public-private partnerships for large investments in secure AI solutions | 0.17 | 38.50 | 0.75 | 43.60 |
| Governmental support for the security of AI labs: Intelligence sharing of relevant threats with the AI labs | 0.14 | 46.75 | 0.67 | 45.07 |
| Secure code generation: R&D on using AI to write more secure code | 0.12 | 41.00 | 0.99 | 44.73 |
| Organizational infrastructure for information/incident sharing: Creating better ways to share information across industry, government, etc.(e.g. ISAC for AI) | 0.08 | 37.89 | 1 | 44.07 |
| AI Control: Frameworks for AI interactions with the environment - develop frameworks for how AI models interact with the outside world that are governable (i.e., how to provide AI models with structured access that is more easily controllable by human-made policies) | 0.03 | 46.00 | 0.75 | 45.74 |
| Secure enclave for weights: develop a high-end, tamper-proof secure enclave for storing and using AI model weights | 0.03 | 35.54 | 0.95 | 43.13 |
| AI Control: Provable prevention of capabilities - research and implement provable security assurances for the prevention of specific actions by unaligned AI (e.g., self-exfiltration) | 0.03 | 35.20 | 0.71 | 42.56 |
| Confidential computing: develop secure and production-ready confidential computing solutions that secure key AI system components (weights, code, etc.) | 0.02 | 42.73 | 0.52 | 43.29 |

## Priority area selection and authorship

The team identified the top ten priority areas according to the importance metric and the top ten priority areas according to the cost-effectiveness metric. Two priority areas appeared in both top-ten lists ("AI security posture assessment" and "Risk management framework for agentic AI"), yielding eighteen unique priority areas. Three thematically close areas were then merged: "Incident reporting" (ranked seventh on cost-effectiveness), "Intelligence sharing of relevant threats with the AI labs" (ranked second on importance), and "Organizational infrastructure for information/incident sharing" (ranked seventh on importance) were combined into a single priority area on information and incident sharing. This merger replaced three priority areas with one, reducing the importance areas from ten to nine, while the cost-effectiveness areas remained ten.

To restore balanced representation across both metrics, the team included the eleventh-ranked priority area. The eleventh-ranked priority area for importance was "Risk management framework for agentic AI," which already appeared in the cost-effectiveness top ten, and therefore introduced no additional priority area. The eleventh-ranked priority area for cost-effectiveness was "Enhanced government background checks for sensitive AI lab personnel," which was not present in the importance ranking, and was accordingly added as a new priority area. This process produced a final set of eighteen priority areas.

To ensure that each priority area reflected specialized knowledge, the team sought authors with demonstrated expertise in the relevant area. Authors were identified through targeted outreach and matched to priority areas based on prior publications, professional experience, and willingness to contribute. Authors were also screened for any potential conflicts of interest.

Authors received standardized guidance for producing a two- to three-page document outlining the priority area, its importance, suggested projects that could be undertaken to address the issue, and elements of success. Irregular and RAND researchers co-authored some priority areas and reviewed all submissions for accuracy, consistency, and clarity.

## Limitations

Several methodological limitations pertain to the study. Despite the authors' best efforts to invite a diverse set of perspectives from across the AI security field, the reliance on professional networks and targeted outreach for interviews and workshop participation may introduce sampling bias and may not fully represent the breadth of viewpoints across the AI security ecosystem. Importance and effort estimates were based on small expert groups, introducing sensitivity to individual judgments even after normalization. The nonlinear importance transformation (–10 to 100) may amplify extreme ratings, and the arithmetic mean may not fully capture variation in smaller samples.

Estimates of existing efforts were constrained by public visibility. Non-public, proprietary, or confidential work, particularly in industry and government, could not be captured, which may lead to undercounting in domains where significant activity is not publicly reported. Effort was made to minimize this limitation through outreach and informal inquiries to identify non-public initiatives, though these measures cannot fully eliminate the gap. Partially mitigating this concern, counts of existing efforts influenced only the cost-effectiveness calculation (and not the importance calculation) and had the least influence among the three factors used in the overall prioritization.

Finally, priority area selection was based directly on the distributions of calculated scores, which may reflect the limitations noted above. Nonetheless, the process was designed to ensure that high-priority areas were consistently represented across the final set.

# Appendix B: Expert ranking tables

**Table B-1. The 10 Most Important Priorities in AI Security**

| Original Task Name | Name of Priority Area in the paper | Cost Effectiveness | Importance | Standard Deviation of Expert Importance Scores* |
|---|---|---|---|---|
| National AI deterrence strategy: USG clarifying strategic posture to deter other nations' theft or sabotage (e.g. committing to sanctions in response to model weight theft) | Develop a National AI Deterrence Strategy | 0.18 | 46.89 | |
| Governmental support for the security of AI labs: Intelligence sharing of relevant threats with the AI labs** | Advance Information and Incident Sharing | 0.14 | 46.75 | |
| AI Control: Frameworks for AI interactions with the environment - develop frameworks for how AI models interact with the outside world that are governable (i.e., how to provide AI models with structured access that is more easily controllable by human-made policies) | Develop Permission Frameworks for AI Agent Interactions | 0.03 | 46.00 | |
| Confidential computing: develop secure and production-ready confidential computing solutions that secure key AI system components (weights, code, etc.) | Implement Confidential Computing to Protect Key AI System Components | 0.02 | 42.73 | |
| Secure code generation: R&D on using AI to write more secure code | Research Secure Code Generation to Write More Secure Code | 0.12 | 41.00 | |
| Governmental support for the security of AI labs: Public-private partnerships for large investments in secure AI solutions | Form Public-Private Partnerships Focused on AI Security | 0.17 | 38.50 | |
| Organizational infrastructure for information/incident sharing: Creating better ways to share information across industry, government, etc.(e.g. ISAC for AI)** | Advance Information and Incident Sharing | 0.08 | 37.89 | |
| AI security posture assessment: develop assessment/auditing | Develop Assessment and Auditing | 1.74 | 36.83 | |
| protocols for AI companies and AI systems | Protocols for AI Companies and AI Systems | | | |
| Secure enclave for weights: develop a high-end, tamper-proof secure enclave for storing and using AI model weights | Develop a Tamper-Proof Secure Weight Module for AI Model Weights | 0.03 | 35.54 | |
| AI Control: Provable prevention of capabilities - research and implement provable security assurances for the prevention of specific actions by unaligned AI (e.g., self-exfiltration) | Establish Formal Verification of Security Properties | 0.03 | 35.20 | |

**Table B-2. The 10 Most Cost-Effective Priority Areas in AI Security**

| Original Task Name | Name of Priority Area in the paper | Cost Effectiveness | Importance | Standard Deviation of Expert Importance Scores* |
|---|---|---|---|---|
| AI Security resources (usable documents and training materials): set up a hub (e.g., public git) of AI Security resources, including materials on defending from nation-states | Establish an AI Security Resource Hub | 2.29 | 8.75 | |
| Elite security red-teaming specifications: create specifications for a top-tier AI security red-teaming effort | Create Specifications for Top-Tier AI Security Red-Teaming | 1.27 | 18.00 | |
| Incident response: develop AI-specific incident response recommendations** | Develop Incident Response Frameworks and Playbooks | 1.00 | 11.50 | |
| Risk management framework for agentic AI (as opposed to RMFs for foundation models) | Develop a Risk Management Framework for Agentic AI | 0.84 | 33.14 | |
| Threat Modeling AI as an attacker: at higher capabilities, the model may attack the infrastructure that is hosting it autonomously (at the direction of an attacker) | Use Threat Modeling to Understand Risks of AI Attacking Its Host Infrastructure | 0.65 | 29.13 | |
| Incident reporting: develop recommendations for protocols for sharing AI security incidents | Advance Information and Incident Sharing | 0.59 | 32.50 | |
| AI security standards: formulate technical standards for securing the critical components of AI models | Establish AI Security Standards | 0.54 | 32.50 | |
| Implications analysis: develop a paradigm of offensive cyber operations to assess which capabilities lead to which societal outcomes | Understand How AI Systems Could Impact Society with Implications Analysis | 0.53 | 13.18 | |
| Taxonomy and language clarification: Improving our taxonomy of "AI security" and how it involves and relates to trustworthiness/control/alignment | Clarify Language Through a Standard Taxonomy | 0.51 | 15.20 | |
| Governmental support for the security of AI labs: Mechanisms for background checks for sensitive personnel*** | Adapt Enhanced Background Checks for Sensitive Personnel | 0.48 | 24.20 | |

***** For the full calculation and the explanation for why we do not include the numerical scores here, see Appendix A.
** Indicates that "Governmental support for the security of AI labs: intelligence sharing of relevant threats with the AI labs," "Incident response: develop AI-specific incident response recommendations," and "Organizational infrastructure for information/incident sharing: creating better ways to share information across industry, government, etc. (e.g., ISAC for AI)" were merged into a single priority area.
*** "Governmental support for the security of AI labs: Mechanisms for background checks for sensitive personnel" was ranked eleventh on the cost-effectiveness, and this table omits "AI security posture assessment: develop assessment/auditing protocols for AI companies and AI systems." For a full explanation of our reasoning, see "Priority area selection and authorship" section in Appendix A.

# Appendix C: RMF table for agentic AI

**Table C-1. Core Components of a Risk Management Framework for Agentic AI**

| RMF Category | Agentic-specific controls / artifacts |
|---|---|
| **Scope and Context Definition** | Define autonomy scope: Establish the intended scope of agent behavior, including system access, permissible tools, and types of decisions the agent may make. Define escalation paths for when human oversight is required. Boundaries could include technical controls such as capability gating, memory limits, and privilege separation. |
| **Risk Identification and Analysis** | Behavioral risk assessment over trajectories: Map how agents generate, revise, and execute plans over time. Identify risks arising from long-run behavior, recursive tool use, delayed feedback, or adaptation to environmental changes. Go beyond output analysis to examine planning processes, goal formation, and the interactions between internal state and external actions. Include mechanisms to detect and limit out-of-distribution behavior, ensuring agents operate within validated contexts and do not generalize beyond tested environmental or task boundaries. |
| **Testing and Validation** | Trajectory testing and evaluation: Develop evaluation environments that simulate high-risk conditions like conflicting goals, adversarial interference, or incomplete information. Assess agent behavior across stateful trajectories and stress-test planning under uncertainty. Use findings to benchmark safe planning patterns and detect failure-prone adaptations. |
| **Risk Control Design** | Capability-threshold controls: Define key capability milestones that trigger additional oversight, such as writing to disk, modifying system configurations, accessing sensitive APIs, or creating sub-agents. Tie governance responses, e.g., audits, reviews, or containment, to these thresholds to ensure that supervision scales with capability. |
| **Monitoring and Detection** | Real-time oversight and response: Implement systems to monitor agent decisions, state transitions, and tool interactions as they occur. Establish mechanisms for automated policy checks, anomaly detection, privilege revocation, and system rollback. Monitoring should be ongoing, not episodic, and tailored to the agent’s autonomy level. |
| **Accountability and Traceability** | Provenance and accountability tracking: Track how agent configurations, permissions, and capabilities were authorized. Maintain logs that map decisions to supervisory contexts, providing post hoc visibility into how incidents developed. Use this information to improve internal accountability, external reporting, and iterative policy updates. |
| **Risk Evaluation and Governance** | Stakeholder-governed risk evaluation: Evaluate whether current and projected agent behavior aligns with organizational goals, ethical principles, and legal obligations. Include stakeholders in defining acceptable risk levels, thresholds for intervention, and transparency standards. Adapt risk posture as the agent’s role or context evolves. |